\documentclass{aa}
\usepackage{natbib}
\usepackage{graphicx}
\usepackage{amssymb}
\usepackage{color}
\usepackage{ulem}

\newcommand{\bl}[1]{\mbox{\boldmath$ #1 $}}

\begin{document}

\title{The early evolution of viscous and self-gravitating circumstellar
disks with a dust component}

\author{Eduard I. Vorobyov\inst{1,2,3}, Vitaly Akimkin\inst{4}, Olga
Stoyanovskaya\inst{5}, Yaroslav Pavlyuchenkov\inst{4}, and Hauyu Baobab Liu\inst{6}}

\institute{Institute of Fluid Mechanics and Heat Transfer, TU Wien, 1060, Vienna, Austria \\
        \and
        Research Institute of Physics, Southern Federal University, Stachki Ave. 194, 344090, 
        Rostov-on-Don, Russia
        \and 
         Department of Astrophysics, University of Vienna, Vienna, 1180, Austria
        \and
        Institute of Astronomy, Russian Academy of Sciences, Pyatnitskaya str. 48, 119017, Moscow, Russia
        \and
        Novosibirsk State University, Lavrentieva str. 2, 630090, Novosibirsk, Russia and \\
        Boreskov Institute of Catalysis, Lavrentieva str. 5, 630090, Novosibirsk, Russia 
        \and
        European Southern Observatory (ESO), Karl-Schwarzschild-Str. 2, D-85748 Garching, Germany
}

\abstract 
{}
{The long-term evolution of a circumstellar disk starting from its
formation and ending in the T~Tauri phase was simulated 
numerically with the purpose of studying the evolution of dust in the disk with distinct values of
viscous $\alpha$-parameter and dust fragmentation velocity $v_{\rm frag}$.} 
{We solved numerical hydrodynamics equations in the thin-disk limit,
which are modified  to include a dust component consisting of two parts:
sub-micron-sized dust and grown dust with a maximum radius
$a_{\rm r}$. The former is strictly coupled to the gas, while the latter
interacts with the gas via friction. The conversion of small to grown
dust, dust growth, and dust self-gravity are also considered.} 
{We found that the process of dust growth known for the older protoplanetary 
phase also holds for the embedded phase of disk evolution.
The dust growth efficiency depends on the radial distance
from the star --- $a_{\rm r}$ is largest in the inner disk and gradually 
declines with radial distance. In the inner disk, $a_{\rm r}$ is limited
by the dust fragmentation  barrier. The process of  small-to-grown dust conversion is very
fast once the disk is formed. The total mass of grown dust  in the disk (beyond 1~AU) 
reaches tens or even hundreds of Earth masses already in the embedded phase of
star formation and even a greater amount of grown dust drifts in the inner,
unresolved 1~AU of the disk.  Dust does not usually grow to radii greater than a few
cm. A notable exception are models with $\alpha\le10^{-3}$, in which case
a zone with reduced mass transport develops in the inner disk and  
dust can grow to meter-sized boulders in the inner 10 AU. 
Grown dust drifts
inward and  accumulates in the inner disk regions. This effect is most pronounced 
in the $\alpha\le10^{-3}$ models where several hundreds of Earth masses can be
accumulated in a narrow region of several AU from the star by the end of embedded phase. 
The efficiency of grown dust accumulation in spiral arms is 
stronger near corotation where the azimuthal velocity of dust grains is closest to the local
velocity of the spiral pattern.
In the framework of the adopted dust growth model, the efficiency of small-to-grown 
dust conversion was found to increase for lower values of $\alpha$ 
and $v_{\rm frag}$.}
{}
\authorrunning{E. Vorobyov et al.}
\titlerunning{Evolution of disks with dust component}
\maketitle

\section{Introduction}
Circumstellar disks hold the key to our understanding of the star and
planet formation process. They serve as a construction site for planets
and as a bridge between collapsing pre-stellar cores and nascent stars.
In the early evolution phase,  properties of circumstellar disks are
controlled by a complex interplay between mass-loading from the parental
core, energy input from the host star, and mass and angular momentum 
transport through the disk.  If the mass infall rate from the core onto
the disk exceeds  the total mass loss rate of the disk (e.g., via 
protostellar accretion, jets, disk winds, etc.), the disk grows in mass
and size and disk gravitational instability (GI) sets in, leading in
some cases to disk fragmentation.

GI and disk fragmentation have manifold implications for the evolution
of circumstellar disks. Disk fragmentation may account for the formation
of giant planets and brown dwarfs, either as companions to host stars or
freely floating objects
\citep{Boss2001,Mayer2002,Boley2009,Nayakshin2010,Nayakshin2017,Vorobyov2013,
Stamatellos2015}. Disk sub-structures formed thanks to GI may serve as
likely locations for dust accumulation and growth \citep{Rice2004,
Nayakshin2010,Gibbons2015} and influence notably the chemical evolution
of young disks \citep{Ilee2011}. Dense gaseous clumps forming in  massive
GI-unstable disks often migrate into the star, causing strong accretion
and luminosity bursts \citep{VB2006,VB2015,Meyer2017}, which in turn may have
important consequences for protostellar disks and collapsing parental
clouds, affecting their gravitational stability, chemical composition,
and dust growth 
\citep{Stamatellos2012,Vorobyov2013b,Jorgensen2015,Frimann2017,Hubbard2017}.

While GI and fragmentation play an important role in the early disk
evolution, especially in the embedded phase, other important phenomena,
such as the magneto-rotational instability (MRI), can also have a
considerable effect on the disk evolution on short and long timescales.
The MRI-induced turbulence can transport mass and angular momentum in the
disk regions where ionization is sufficiently high
\citep[e.g.][]{Turner2014}, and induce strong accretion and luminosity
bursts in the inner  disk \citep[e.g][]{Zhu2009,Ohtani2014}. However,
computing disk evolution with the  self-consistently induced and
sustained MRI turbulence is a challenging  task requiring non-ideal 
magnetohydrodynamics simulations coupled with the stellar radiation input, dust evolution effects, 
and ionization balance calculations \citep{2015ARep...59..747A,2016ApJ...833...92I}.
Such models enable disk simulations only for a
limited evolution period  and for a narrow set of circumstellar disk
parameters \citep[e.g.][]{Flock2012,Simon2015}.  Therefore, numerical
simulations  adopting the $\alpha$-prescription of \citet{SS1973}  to
mimic the effect of turbulent viscosity have been  routinely employed to
study the long-term evolution of circumstellar disks 
\citep[e.g.][]{VB2009,Vorobyov2010,Visser2010,Rice2010,Kimura2016}. 
These simulations have shown the importance of 
disk self-gravity and viscosity for the evolution and global properties
(disk masses, radii) of circumstellar disks. 

In this paper, we continue our efforts to study the long-term evolution
of self-gravitating and viscous circumstellar disks started in
\citet{VB2009}.  We now employ a numerical hydrodynamics code that, in
addition to the gaseous component, also  includes the dust component. 
%The evolution of the central star is described by the
%Lyon stellar evolution code \citep{Chabrier97,Baraffe2009}. 
The evolution of the dust component includes the conversion of small grains into larger
grains due to coagulation (using the monodisperse growth approach as in
\cite{1997A&A...319.1007S}) and drift of the grown dust relative to the
gas.  Similar modeling of the mutual gas and dust evolution was recently
presented in \citet{Gonzalez2017}. Our study extends these
theoretical efforts by accounting for the self-consistent formation of a 
circumstellar disk from a parental cloud and by considering the
accompanying effect of disk gravitational instability.
In this study, we consider the case of a constant $\alpha$-parameter. 
Unlike many previous  studies of  self-gravitating disks with a dust component 
\citep{Rice2004,Cha2011,Gibbons2012,Booth2016}, we follow the disk
evolution starting from its formation from a collapsing pre-stellar core
and ending in the early T~Tauri phase when essentially all of the 
collapsing core has
dissipated. We pay specific attention to the resulting  properties of the
gaseous and dusty components of the circumstellar disk, including the
efficiency of dust growth, inward drift, and accumulation.  
Our present development is immediately relevant, since the previous observational 
studies of a sample of Class 0 young stellar objects 
have suggested that grain growth begins to significantly change the values of dust 
opacity spectral index ($\beta$) on 10$^{2}$ AU scales at the $\sim$1--10~mm 
wavelength range from late Class~0 to early Class~II stages 
\citep{Li2017}. In the later stages of disk evolution, observations indicate dust-to-gas 
ratios that are higher than the ISM value of 0.01, also signifying strong evolution in the disk
dust composition \citep[e.g.][]{Williams2014,Ansdell2016}.

The paper is organized as follows. In Sect.~\ref{diskmodel}, we present
the detailed description of our numerical model. The main results are
presented in Sect.~\ref{results}. The parameter space study is conducted in
Sect.~\ref{paramspace}. The model caveats and future
improvements are discussed in Sect.~\ref{caveats}. The main results are
summarized in Sect.~\ref{summary}. Several essential test problems
addressing the performance of our numerical scheme on the dust component
are shown in  Appendix~\ref{Appendix}.

\section{Protostellar disk model}
\label{diskmodel}

The numerical model for the formation and evolution of a star and its
circumstellar disk (FEoSaD) is described in detail in \citet{VB2015} and
\citet{Dong2016}. Here, we briefly review its main constituent parts and
describe new additions and updates introduced to model the co-evolution
of a dusty disk. Numerical simulations start from a collapsing
pre-stellar core of a certain mass, angular momentum, temperature, and
dust-to-gas ratio. The properties of the nascent star are calculated
using the stellar evolution tracks derived using the STELLAR evolution code 
\citep{Yorke2008,Vorobyov2017}, while the formation and
long-term evolution of the gaseous and dusty disk components are
described using numerical hydrodynamics simulations in the two-dimensional ($r,\phi$) thin-disk
limit.  The evolution of the star and disk are interconnected: the
star grows according to the mass accretion rate
provided by hydrodynamic simulations and heats the disk according to its photospheric 
and accretion luminosities.

\subsection{Gaseous component}
\label{gaseous}

The main physical processes taken into account when modeling the disk
formation and evolution include viscous and shock heating, irradiation by
the forming star,  background irradiation with a uniform temperature of $T_\mathrm{bg}=20$\,K 
set equal to the initial temperature of the natal cloud core,
radiative cooling from the disk surface, friction between the gas and
dust components, and self-gravity of gaseous and dusty disks.  The code
is written in the thin-disk limit, complemented by a calculation of the
gas vertical  scale height using an assumption of local hydrostatic
equilibrium as described in \citet{VB2009}. The resulting  model has a
flared structure (because the disk vertical scale height increases with
radius), which guaranties that both the disk and envelope receive a
fraction of the irradiation energy  from the central protostar. The
pertinent equations of mass, momentum, and energy transport for the gas
component are
\begin{equation}
\label{cont}
\frac{{\partial \Sigma_{\rm g} }}{{\partial t}}   + \nabla_p  \cdot 
\left( \Sigma_{\rm g} \bl{v}_p \right) =0,  
\end{equation}
\begin{eqnarray}
\label{mom}
\frac{\partial}{\partial t} \left( \Sigma_{\rm g} \bl{v}_p \right) +  [\nabla \cdot \left( \Sigma_{\rm
g} \bl{v}_p \otimes \bl{v}_p \right)]_p & =&   - \nabla_p {\cal P}  + \Sigma_{\rm g} \, \bl{g}_p + \nonumber
\\ 
&+& (\nabla \cdot \mathbf{\Pi})_p,
\end{eqnarray}
\begin{equation}
\frac{\partial e}{\partial t} +\nabla_p \cdot \left( e \bl{v}_p \right) = -{\cal P} 
(\nabla_p \cdot \bl{v}_{p}) -\Lambda +\Gamma + 
\left(\nabla \bl{v}\right)_{pp^\prime}:\Pi_{pp^\prime}, 
\label{energ}
\end{equation}
where subscripts $p$ and $p^\prime$ refer to the planar components
$(r,\phi)$  in polar coordinates, $\Sigma_{\rm g}$ is the gas mass
surface density,  $e$ the internal energy per surface area,  ${\cal P}$
the vertically integrated gas pressure calculated via the ideal  equation
of state as ${\cal P}=(\gamma-1) e$ with $\gamma=7/5$, $\bl{v}_{p}=v_r
\hat{\bl r}+ v_\phi \hat{\bl \phi}$  the gas velocity in the disk plane,
%$\bl{f}$ the friction force per unit mass between the dust and gas disk
%components,   
and $\nabla_p=\hat{\bl r} \partial / \partial r + \hat{\bl
\phi} r^{-1} \partial / \partial \phi $ the gradient along the planar
coordinates of the disk.  The gravitational acceleration in the disk
plane,  $\bl{g}_{p}=g_r \hat{\bl r} +g_\phi \hat{\bl \phi}$, takes into
account self-gravity of the gaseous and dusty disk components found by
solving for the Poisson integral \citep[see details in][]{VB2010} and the
gravity of the central protostar when formed. Turbulent viscosity is
taken into account via the viscous stress tensor  $\mathbf{\Pi}$, the
expression for which can be found in \citet{VB2010}. We parameterize the
magnitude of kinematic viscosity $\nu=\alpha c_{\rm s} H_{\rm g}$  using
the alpha prescription of \citet{SS1973} with a constant
$\alpha$-parameter, where $c_{\rm s}$ is the sound speed of gas and
$H_{\rm g}$ is the gas vertical scale height.

The cooling and heating rates $\Lambda$ and $\Gamma$ take the disk
blackbody cooling and heating due to stellar and background irradiation
into account based on the analytical solution of the radiation transfer
equations in the vertical  direction
\citep[see][for detail]{Dong2016}\footnote{The cooling and heating rates in \citet{Dong2016}
are written for one side of the disk and need to be multiplied by a factor of 2.}:
\begin{equation}
\Lambda=\frac{8\tau_{\rm P} \sigma T_{\rm mp}^4 }{1+2\tau_{\rm P} + 
{3 \over 2}\tau_{\rm R}\tau_{\rm P}},
\end{equation}
where $T_{\rm mp}={\cal P} \mu / {\cal R} \Sigma_{\rm g}$ is the midplane
temperature,  $\mu=2.33$ the mean molecular weight,  $\cal R$ the
universal  gas constant, $\sigma$ the Stefan-Boltzmann constant, 
{$\tau_{\rm R}=\kappa_{\rm R} \Sigma_{\rm d, sm}/2$  and $\tau_{\rm
P}=\kappa_{\rm P} \Sigma_{\rm d,sm} / 2$ the  Rosseland and Planck
optical depths to the disk midplane, which are calculated from the
evolving surface density of small dust population $\Sigma_{\rm d,sm}$
(see Sect.~\ref{dustycomp}), and  $\kappa_P$, and $\kappa_R$, g~cm$^{-2}$
are the Planck and Rosseland mean opacities taken from
\citet{Semenov2003}}.

The heating function per surface are of the disk is expressed as
\begin{equation}
\Gamma=\frac{8\tau_{\rm P} \sigma T_{\rm irr}^4 }{1+2\tau_{\rm P} + {3 \over 2}\tau_{\rm R}\tau_{\rm
P}},
\end{equation}
where $T_{\rm irr}$ is the irradiation temperature at the disk surface 
determined by the stellar and background black-body irradiation as
\begin{equation}
T_{\rm irr}^4=T_{\rm bg}^4+\frac{F_{\rm irr}(r)}{\sigma},
\label{fluxCS}
\end{equation}
where $F_{\rm
irr}(r)$ is the radiation flux (energy per unit time per unit surface
area)  absorbed by the disk surface at radial distance  $r$ from the
central star. The latter quantity is calculated as 
\begin{equation}
F_{\rm irr}(r)= \frac{L_\ast}{4\pi r^2} \cos{\gamma_{\rm irr}},
\label{fluxF}
\end{equation}
where $\gamma_{\rm irr}$ is the incidence angle of radiation arriving at
the disk surface (with respect to the normal) at radial distance $r$. The
incidence angle is calculated using a flaring disk surface as described
in \citet{VB2010}. The stellar luminosity $L_\ast$ is the sum of the
accretion luminosity  $L_{\rm \ast,accr}=(1-\epsilon) G M_\ast \dot{M}/2
R_\ast$ arising from the gravitational energy of accreted gas and the
photospheric luminosity $L_{\rm \ast,ph}$ due to gravitational
compression and deuterium burning in the stellar interior. The stellar
mass $M_\ast$ and accretion rate onto the star $\dot{M}$ are determined
using the amount of gas passing through the sink cell (see Sect.~\ref{Solution}). 
The properties of
the forming protostar ($L_{\rm \ast,ph}$ and radius $R_\ast$) are
calculated using the stellar evolution tracks derived using the STELLAR code.
The fraction of
accretion energy absorbed by the star $\epsilon$ is set to 0.1.

\subsection{Dusty component}
\label{dustycomp}
In our model, dust consists of two components: small micron-sized dust
and grown dust.  The former constitutes the initial reservoir for dust
mass and provides the  main input to opacity and the latter allows us to
study dust growth and drift.  Small dust is assumed to be coupled to
gas, meaning that we only solve the continuity equation for small dust
grains, while the dynamics of grown dust is controlled by friction with
the gas component and by the total gravitational potential of the star,
gaseous and dusty components. Small dust can turn into grown dust and
this process is considered by calculating the dust growth rate and the
maximum radius of grown dust. The resulting continuity and
momentum equations for small and grown dust are as follows
\begin{equation}
\label{contDsmall}
\frac{{\partial \Sigma_{\rm d,sm} }}{{\partial t}}  + \nabla_p  \cdot 
\left( \Sigma_{\rm d,sm} \bl{v}_p \right) = - S(a_{\rm r}),  
\end{equation}
\begin{equation}
\label{contDlarge}
\frac{{\partial \Sigma_{\rm d,gr} }}{{\partial t}}  + \nabla_p  \cdot 
\left( \Sigma_{\rm d,gr} \bl{u}_p \right) = S(a_{\rm r}),  
\end{equation}
\begin{eqnarray}
\label{momDlarge}
\frac{\partial}{\partial t} \left( \Sigma_{\rm d,gr} \bl{u}_p \right) +  [\nabla \cdot \left( \Sigma_{\rm
d,gr} \bl{u}_p \otimes \bl{u}_p \right)]_p  &=&   \Sigma_{\rm d,gr} \, \bl{g}_p + \nonumber \\
 + \Sigma_{\rm d,gr} \bl{f}_p + S(a_{\rm r}) \bl{v}_p,
\end{eqnarray}
where $\Sigma_{\rm d,sm}$ and $\Sigma_{\rm d,gr}$ are the surface
densities of small and grown dust, $\bl{u}_p$ the planar components of
the grown dust velocity, $S(a_{\rm r})$ the rate of dust growth per unit
surface area, $\bl{f}_p$ the drag force per unit mass between dust and gas, 
and $a_{\rm r}$ the maximum radius of grown dust. 
To derive the expression for $S(a_{\rm r})$, we assume a power-law
distribution of dust grains over radius $n(a)= C a^{-p}$  
and note that the total dust mass in a specific grid cell does not change due to dust
growth so that
%\begin{eqnarray}
%{4 \pi \rho_{\rm s} \over 3} C^n \int \limits_{a_{\rm min}} \limits^{a^n_{\rm r}} a^{3-p} da
%&=& {4 \pi \rho_{\rm s} \over 3} C^{n+1} \int \limits_{a_{\rm min}} \limits^{a^{n+1}_{\rm r}} 
%a^{3-p} da \nonumber \\ 
%&=&  \Sigma_{\rm d,tot} \Delta S,
%\end{eqnarray}
\begin{equation}
{4 \pi \rho_{\rm s} \over 3} C^n \int \limits_{a_{\rm min}} \limits^{a^n_{\rm r}} a^{3-p} da
= {4 \pi \rho_{\rm s} \over 3} C^{n+1} \int \limits_{a_{\rm min}} \limits^{a^{n+1}_{\rm r}} 
a^{3-p} da =  \Sigma_{\rm d,tot} \Delta S,
\end{equation}
where $C^n$ and $C^{n+1}$ are the normalization constants, $\Sigma_{\rm d,tot}=\Sigma_{\rm d,gr}+
\Sigma_{\rm d,sm}$ the total surface density of dust, $\Delta S$ the
surface area of a specific grid cell, indices $a_{\rm r}^n$ and $a_{\rm
r}^{n+1}$ the maximum dust radii at the current and next time
steps, $a_{\rm min}=0.005~\mu m$ the minimum radius of small dust grains,
and $p=3.5$ the slope of dust distribution over radius.

\begin{figure}
\begin{centering}
\resizebox{\hsize}{!}{\includegraphics{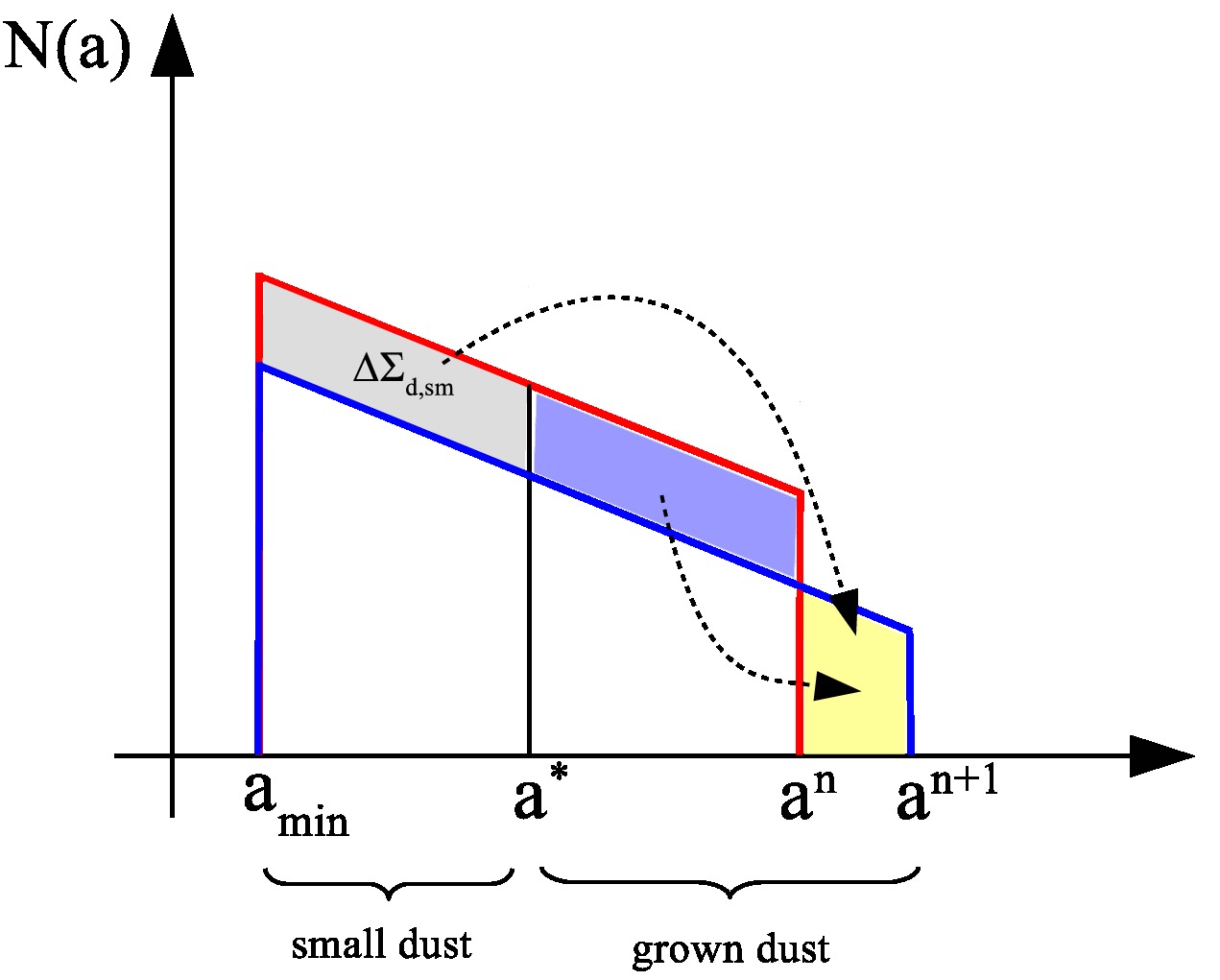}}
\par\end{centering}
\centering{}\protect\protect\protect\caption{\label{fig0} 
{Illustration of the adopted scheme for dust growth. 
The size distributions of dust grains at the current $n$
and next $n+1$ time steps are shown with the red and blue lines,
respectively. Here, $a_{\rm min}$ is the minimum radius of small dust grains,
while $a_{\rm r}^{n}$ and $a_{\rm r}^{n+1}$ are the maximum
radii of the grown dust at the current and next time step.  
Both distributions are assumed to have the same power
law index. The grey area represents the amount of small dust
$\Delta \Sigma_{\rm d,sm}$ (per surface area) converted to grown dust during
one hydrodynamic time step $\Delta t$.} The blue area between $a_\ast$ and $a_n$
is also transferred above $a_n$, but this operation does not change the total mass of grown dust.}
\end{figure}

Figure~\ref{fig0} illustrates our dust growth scheme showing the dust
distribution at the current and next time steps $n$ and $n+1$ with the
red and blue lines, respectively. Here, $a_\ast=1.0~\mu m$ is a
threshold value between small and grown dust components. The grey
area represents schematically the amount of small dust 
$\Delta \Sigma_{\rm d,sm}$ (per surface area)
converted to grown dust during one hydrodynamic time step $\Delta t$.
We note that the area confined between
$a_\ast$ and $a_{\rm r}^n$ (highlighted by the blue color)  should also be transferred 
above $a_{\rm r}^n$, but it does not change the mass of grown dust.
% and is therefore not taken into account in the subsequent analysis.} 
The corresponding mass per surface area in a specific numerical cell
can be expressed as
%\begin{eqnarray}
%\Delta \Sigma_{\rm d,sm} &=& \Sigma_{\rm d,sm}^{n+1}- \Sigma_{\rm d,sm}^n = \nonumber \\ 
%&-&\Sigma_{\rm d,tot}^n  
%{ \int \limits_{a_{\rm r}^n} \limits^{a_{\rm r}^{n+1}} a^{3-p} da \int \limits_{a_{\rm min}} 
%\limits^{a_\ast} a^{3-p} da \over \int \limits_{a_{\rm min}} \limits^{a_{\rm r}^n} a^{3-p} da 
%\int \limits_{a_{\rm min}} \limits^{a_{\rm r}^{n+1}} a^{3-p} da   },
%\label{DustGrowth}
%\end{eqnarray}
\begin{equation}
\Delta \Sigma_{\rm d,sm} = \Sigma_{\rm d,sm}^{n+1}- \Sigma_{\rm d,sm}^n =
-\Sigma_{\rm d,tot}^n  
{ \int \limits_{a_{\rm r}^n} \limits^{a_{\rm r}^{n+1}} a^{3-p} da \int \limits_{a_{\rm min}} 
\limits^{a_\ast} a^{3-p} da \over \int \limits_{a_{\rm min}} \limits^{a_{\rm r}^n} a^{3-p} da 
\int \limits_{a_{\rm min}} \limits^{a_{\rm r}^{n+1}} a^{3-p} da   },
\label{DustGrowth}
\end{equation}
and the rate of dust growth $S(a_{\rm r})$ can finally be expressed as
\begin{equation}
\label{GrowthRate}
S(a_{\rm r}) = - {\Delta \Sigma_{\rm d,sm} \over \Delta t  }.
\end{equation}
The minus sign reflects the fact that the difference 
$\Delta \Sigma_{\rm d,sm}=\Sigma_{\rm d,sm}^{n+1}-\Sigma_{\rm d,sm}^n$ becomes negative
when dust grows in a specific cell, i.e., when $a_{\rm r}^{n+1}>a_{\rm r}^n$.
%meaning that the amount of small dust in a specific cell
%decreases. At the same time, the amount of grown dust increases correspondingly, which 
%is reflected by the minus sign before $S(a_{\rm r})$ in Equation~(\ref{GrowthRate}).  }
We note that $S(a_{\rm r})$ is non-zero only if $a_{\rm r}>a_\ast$.  In
our models, small dust has radii in the ($a_{\rm min}:a_\ast$) range, 
while grown dust has radii in the ($a_\ast:a_{\rm r}$) range with the
temporally and spatially varying maximum radius of grown dust
$a_{\rm r}$. Equation~(\ref{DustGrowth}) also implicitly allows for
conversion of grown dust to small dust due to collisional fragmentation
if $a_{\rm r}^{n+1}<a_{\rm r}^n$.
This can occur when grown grains are advected in the disk regions where the 
fragmentation barrier set by Equation~(\ref{afrag}) is lower than the current size of the advected particles.
In this case, the radius of dust particles is reduced to match the fragmentation barrier, which formally corresponds to $a_{\rm r}^{n+1}<a_{\rm r}^n$ and to fragmentation of grown dust back to small dust.
Finally, we note that the adopted dust size distribution $n(a)=Ca^{-p}$ 
is used only to evaluate $S(a_{\rm r})$ using the total dust density in 
a specific cell. The individual densities of small and grown dust can also change due to
advection and this may produce a discontinuity at $a_\ast$, which is not taken into account
in the current model.

It is worthwhile to analyze how the value of $\Delta \Sigma_{\rm d,sm}$ depends
on $a_{\rm r}$. To do this, we express $\Delta \Sigma_{\rm d,sm}$ as
\begin{equation}
 \Delta \Sigma_{\rm d,sm} = - \Sigma_{\rm d,tot}
\frac{I_1 I_3}{I_2(I_2+I_3)},
\label{fullInt}
\end{equation}
 where
\begin{equation}
 I_1=\int \limits_{a_{\rm min}} \limits^{a_\ast} a^{3-p} da, \hspace{0.3cm}
 I_2=\int \limits_{a_{\rm min}}      \limits^{a_{\rm r}^{n}} a^{3-p} da, \hspace{0.3cm}
 I_3=\int \limits_{a_{\rm r}^n} \limits^{a_{\rm r}^{n+1}} a^{3-p} da.
\end{equation}
We note that $I_3$ is usually much smaller than $I_2$,
because the difference $\Delta a_{\rm r} = {a_{\rm r}^{n+1}} - {a_{\rm r}^{n}}$
caused by dust growth  during one time step is also small. Under these assumptions,
we can write:
\begin{equation}
\Delta \Sigma_{\rm d,sm} \approx - \Sigma_{\rm d,tot} \frac{I_1 I_3}{I_2^2}.
\end{equation}
Noting further that $a_{\rm r}^n \gg a_{\rm min}$ and making an additional assumption that
$p<4$, the integral $I_2$ can be written as:
\begin{equation}
 I_2=\frac{1}{4-p} \left( (a_{\rm r}^{n})^{4-p} - a_{\rm min}^{4-p} \right)
 \approx \frac{1}{4-p} (a_{\rm r}^{n})^{4-p}.
\end{equation}
Furthermore, because  ${a_{\rm r}^{n+1}} - {a_{\rm r}^{n}}$ is small, $I_3$ can be \
approximated by the following expression:
\begin{equation}
I_3 \approx (a_{\rm r}^{n})^{3-p} \Delta a_{\rm r}.
\end{equation}
Noting  that $I_1$ is constant, we finally obtain:
\begin{equation}
 \Delta \Sigma_{\rm d,sm} \propto - \Sigma_{\rm d,tot}\,
(a_{\rm r}^n)^{p-5}\, \Delta a_{\rm r}.
 \label{inverse2}
\end{equation}
We verified by numerical integration that Equation~(\ref{fullInt})
and the approximate formula~(\ref{inverse2}) yield similar results
in all the adopted ranges of dust sizes.
Therefore, for the adopted power law $p=3.5$, $\Delta \Sigma_{\rm d,sm}\propto 
a_{\rm r}^{-1.5}$, meaning that the conversion of small to grown dust is more
efficient when $a_{\rm r}$ is small.

The evolution of the maximum radius $a_{\rm r}$ is described by the
advection equation  with a non-zero source term:
\begin{equation}
{\partial a_{\rm r} \over \partial t} + (u_{\rm p} \cdot \nabla_p ) a_{\rm r} = \cal{D}.
\label{dustA}
\end{equation}
where the growth rate $\cal{D}$ accounts for the dust evolution due to
coagulation and fragmentation. We note that when $\cal{D}$ is zero,
Equation~(\ref{dustA}) turns into the Lagrangian or co-moving equation
for $a_{\rm r}$, guaranteeing that the bulk motion of grown dust, such as
compression or rarefaction, does not change the value of $a_{\rm r}$ in
the absence of dust growth.  We checked this essential property of
Equation~(\ref{dustA}) by initiating a gravitational collapse of a cloud
core without dust growth and confirmed that $a_{\rm r}$ stays constant
and equal to its initial value as the core collapses and disk forms and
evolves.

We write the source term $\cal{D}$ as
\begin{equation}
\cal{D}=\frac{\rho_{\rm d} {\it v}_{\rm rel}}{\rho_{\rm s}},
\label{GrowthRateD}
\end{equation}
where $\rho_{\rm d}$ is the total dust volume density, ${\rm v}_{\rm rel}$ the
dust-to-dust collision  velocity,  and  $\rho_{\rm s}=2.24$\,g~cm$^{-3}$
the solid (internal) density of grains. 
The adopted approach is an extension of the monodisperse model of 
\citet{1997A&A...319.1007S}, so that we use the total dust volume density 
rather than that of grown dust only.
This allows us to account for collisions not only within the grown dust ensemble, but also 
between the grown and small dust grains. Because $v_{\rm rel}$ refers to the grown 
dust, the source term $\cal D$ may be 
underestimated at the initial stages of dust evolution when the collision velocities are 
dominated by the fast  Brownian motion of small dust grains. However, at the later stages, when 
the turbulence-induced velocities start to dominate over the Brownian motion, the 
small difference in $v_{\rm rel}$ 
for grown-to-grown and grown-to-small dust grain collisions  
\citep[see fig.~3 in][]{2007A&A...466..413O} justify the adopted 
approach. Because  the basic dust dynamics
is described in terms of the dust surface density, we use the following expression
\begin{equation}
 \rho_{\rm d}=\frac{\Sigma_{\rm d,tot}}{\sqrt{2\pi}\,{H_{\rm d}}}
\end{equation}
as a proxy for the midplane dust volume density. We note that
here we assumed that the scale heights of both small and grown dust are similar,
and are equal to the scale height of the grown dust $H_{\rm d}$.
This assumption can be violated in
the presence of efficient dust settling, when small grains are expected to be
well mixed with the gas (thus having the same scale height with the gas), but
the scale height of the grown dust is expected to be smaller than that of the gas. 
This effect can decrease the total volume density of dust and the growth rate efficiency.
However, as we demonstrate in Sect.~\ref{paramspace}, dust settling has a minor effect on dust growth
as compared to the effects of turbulent viscosity and dust fragmentation barrier.
The dust vertical scale
height {$H_{\rm d}=H_{\rm d}(a_{\rm r})$} is calculated adopting the approach of
\citet{2001A&A...378..180K} (their~Eq.~(10)), which links the gas and
dust vertical scale heights taking the dust sedimentation into account.

The dust-to-dust collision velocity is calculated as $v_{\rm rel}=(v_{\rm
th}^2+v_{\rm turb}^2)^{1/2}$. For the main sources of
relative velocities between the dust grains, we consider the Brownian
motions with dust collision velocities
\begin{equation}
 v_{\rm th}=\sqrt{\frac{16 k_{\rm B} T_{\rm mp}}{\pi m_{\rm a}}},
\end{equation}
and the turbulence-induced velocities \citep{2007A&A...466..413O, 2012A&A...539A.148B}
\begin{equation}\label{ggrelturb}
 v_{\rm turb}=\sqrt{3\,\alpha\,{\rm St}}\,c_{\rm s},
\end{equation}   
where  $k_{\rm B}$ is the Boltzmann constant, $T_{\rm mp}={\cal P} \mu /
{\cal R} \Sigma$ the gas midplane temperature, and $m_{\rm a}$ the mass
of grown dust grains with maximum radius $a_{\rm r}$. The Stokes number
is defined as 
\begin{equation}
 {\rm St}=\frac{\Omega_{\rm K}\rho_{\rm s} a_{\rm r}}{\rho_{\rm g}c_{\rm s}},
 \label{StokesN}
\end{equation}
where the gas volume density is calculated as $\rho_{\rm g}=\Sigma_{\rm
g}/(\sqrt{2\pi} {H_{\rm g}}) $.
To summarize the adopted dust growth mechanism, we first calculate the growth rate
$\cal{D}$ from Equation~(\ref{GrowthRateD}) using the local quantities in a specific cell.
The resulting value of $\cal{D}$ is used to update the value of $a_r$ using Equation~(\ref{dustA}),
which in turn is used in Equations~(\ref{GrowthRate}) and (\ref{DustGrowth}) to compute the amount 
of small dust converted to grown dust and the growth rate $S(a_{\rm r})$.

To simulate the fragmentation of dust grains due to mutual collisions, we
assume that the maximum radius of a dust grain can not exceed the
fragmentation  barrier~\citep{2012A&A...539A.148B} defined as follows
\begin{equation}
 a_{\rm frag}=\frac{2\Sigma_{\rm g}v_{\rm frag}^2}{3\pi\rho_{\rm s}\alpha c_{\rm s}^2},
 \label{afrag}
\end{equation}
where  $v_{\rm frag}$ is a threshold value for the fragmentation
velocity.  This effectively means that whenever $a_r$ exceeds $a_{\rm frag}$, the 
growth rate $\cal{D}$ is set to zero. We note that even if ${\cal D}=0$ the dust size in a given cell
can change due to advection, as Equation~(\ref{dustA}) implies.
%We set $v_{\rm frag}$  to $30$~m~s$^{-1}$. 
The fragmentation velocities of dusty aggregates 
depend on their properties and may vary in wide limits, typically $\sim 1-10$~m~s$^{-1}$ 
\citep{Dominik1997,Benz2000,Blum2008}. 
The numerical simulations of collisions between icy dust aggregates 
suggest $v_{\rm frag}\approx20$~m~s$^{-1}$ for sintered aggregates \citep{Sirono2014} and 
$v_{\rm frag}\approx50$~m~s$^{-1}$ for non-sintered aggregates \citep{Wada2009}. 
We adopt $v_{\rm frag}=30$~m~s$^{-1}$ in the fiducial model presented in Sect.~\ref{results} 
and explore lower values of $v_{\rm frag}$ in Sect.~\ref{paramspace}. In future studies, we plan to
adopt a more detailed approach, which will consider the dependence of $v_{\rm frag}$ on 
the grain properties, such as their size, ice content, and porosity.

The comparison of the monodisperse growth model of \citet{1997A&A...319.1007S} with the
full-fledged dust growth model was  conducted in~\citet{2012A&A...539A.148B},
showing the validity of the monodisperse approach. 
However, we want to point out two possible caveats with
this approximation. First, grains that would experience a succession of growth and fragmentation events during their evolution, rather than directly growing to a fragmentation boundary, would reach 
the limiting size after a longer time.  Second, the study of \citet{2012A&A...539A.148B} applied
to older disks (than in the current work), which usually evolve slower than the dust size distribution.
Here, we study the early, more dynamical stage of disk evolution, and it is not obvious that the limiting size approximation would also match a more detailed dust evolution 
since the disk structure evolves much faster. More accurate models of dust evolution need to be 
employed in the future to test our assumptions. 
Finally, we note that \citet{Pinilla2016} and
\citet{Gonzalez2017} also applied models for calculating the
dust evolution in circumstellar disks, which are different in certain aspects. 
More specifically, the model of \citet{Pinilla2016} uses the full-fledged 
dust evolution model of \citet{Birnstiel2010}, and not the two-population approximation. 
\citet{Gonzalez2017} use the monodisperse growth model of \citet{1997A&A...319.1007S} and 
added fragmentation as a decrease in size each time the relative velocities of grains exceed 
the fragmentation threshold, instead of limiting the growth to a grain size depending on that 
threshold.

\subsection{Drag force}
We express the drag force per unit mass between dust and gas following
the common practice \citep[e.g.][]{Rice2004,Cha2011,Zhu2012} as
\begin{equation}
\bl{f}_p= { \bl{v}_p - \bl{u}_p \over t_{\rm stop}},
\label{force}
\end{equation}
where $t_{\rm stop}$ is the stopping time expressed in the Epstein regime as
\begin{equation}
t_{\rm stop} = { a_{\rm r} \rho_{\rm s} H_{\rm g} \sqrt{2 \pi} \over \Sigma_{\rm g} c_{\rm s} }.
\label{tstop}
\end{equation}
Since this approach is valid only for the Epstein drag, we limit the
growth of dust in our modeling to a radius $a_{\rm r}<9\lambda/4$ by manually
setting $\cal{D}$ to zero if this condition is
violated\footnote{Equation~(\ref{force}) can in principle be used in the
Stokes regime for as long as the Reynolds number $Re=4a |\bl{u}_p -
\bl{v}_p| / (\lambda c_{\rm s})$ is smaller than unity. In this case,
$t_{\rm stop}$ needs to be multiplied by $4a/(9\lambda)$. 
 However, the Stokes regime with $Re<1$  is usually very narrow.}. Here,
$\lambda$ is the free mean path of grown dust particles defined as
$\lambda= m_{\rm H2} \sqrt{2\pi} H_{\rm g}/(7\times10^{-16} \Sigma_{\rm
g})$, where $m_{\rm H2}$ is the mass of hydrogen molecule
\citep{Rice2004}. We note that our definition of $a_{\rm r}$ in Equation~(\ref{tstop}) 
implies that we follow the 
dynamics of dust grains of maximum size, whereas in reality there is a spectrum of grown dust grains
between $a_\ast$ and $a_{\rm r}$. An alternative would be to define $a_{\rm r}$ as a weighted 
average between $a_\ast$ and $a_{\rm r}$, in which case the dynamics of an averaged population of 
grown dust grains would be computed. We defer these and other sophistications of our model  to follow-up
studies. 
%In this work, we therefore focus on the dynamics of dust grains of maximum size an }

\subsection{Solution procedure and boundary conditions}
\label{Solution}
Equations (\ref{cont})--(\ref{energ}) and
(\ref{contDsmall})--(\ref{momDlarge})  and (\ref{dustA}) are solved using  the operator-split
solution procedure as described in the ZEUS-2D code \citep{SN1992}. The
solution is split in the transport and source steps. In the transport
step, the update of hydrodynamic quantities due to advection is done
using the third-order piecewise parabolic interpolation scheme of
\citet{CW1984}. This step also considers the change of maximum
dust radius due to advection. We note that the term $(\bl{u}_p\cdot
{\nabla}_p) a$ on the r.h.s. of  Equation~(\ref{dustA}) is not  exactly
advection, but the derivative along the direction of dust velocity
$\bl{u}_p$. However, this term can be recast as advection
${\nabla}_p\cdot (a \bl{u}_p)$  minus  a correction term $a ({\nabla}_p
\cdot \bl{u}_p)$, the latter applied to the solution procedure at the
advection step. We finally note that we use the FARGO algorithm to ease
the strict limitations on the Courant condition, which occur in numerical
simulations of Keplerian disks using  the curvilinear coordinate systems
converging towards the origin \citep{Masset2000}.

In the source step, the update of hydrodynamic quantities due to gravity,
viscosity, cooling and heating,  and also friction between gas and dust
components is performed. This step also considers the transformation of
small to grown dust and also the increase in dust radius $a_{\rm r}$ due
to growth (Equations~\ref{DustGrowth}--\ref{dustA}). To account for the
friction terms, we apply the semi-implicit scheme of \citet{Cha2011} in a
modified  source step defined as
\begin{equation}
{\bl{u}_p^{n+1} - \bl{u}_p^n \over \Delta t } = {\bl{v}_p^n - \bl{u}_p^{n+1} \over t_{\rm stop}} 
+ \tilde{\bl{g}}, 
\end{equation}
where $\tilde{\bl{g}}$ includes all non-friction terms on the r.h.s. of
Equation~(\ref{momDlarge}) and $\Delta t$ is the hydrodynamic time step.
This equation can be recast in the following form
\begin{equation}
\label{eq:dustScheme}
\bl{u}_p^{n+1}={t_{\rm stop} \over \Delta t +t_{\rm stop}} \left( {\Delta t \over t_{\rm stop}} \bl{v}_p^n  + \bl{u}_p^n + \Delta t \, \tilde{\bl{g}} \right) .
\end{equation}
This form correctly reproduces the two limiting cases: $\bl{u}_p^{n+1} =
\bl{v}_p^{n}$  for $t_{\rm stop} \rightarrow 0$ and $\bl{u}_p^{n+1} =
\bl{u}_p^n+\Delta t \, \tilde{\bl{g}}$ for $t_{\rm stop} \rightarrow
\infty$.

The update of the internal energy per surface area due to cooling
$\Lambda$ and  heating $\Gamma$  is done implicitly using the
Newton-Raphson method of root finding, complemented by the bisection
method where the Newton-Raphson iterations fail to converge. The implicit
solution is applied to avoid too small time steps that may emerge in
regions of fast heating or cooling.

\begin{figure}
\begin{centering}
\resizebox{\hsize}{!}{\includegraphics{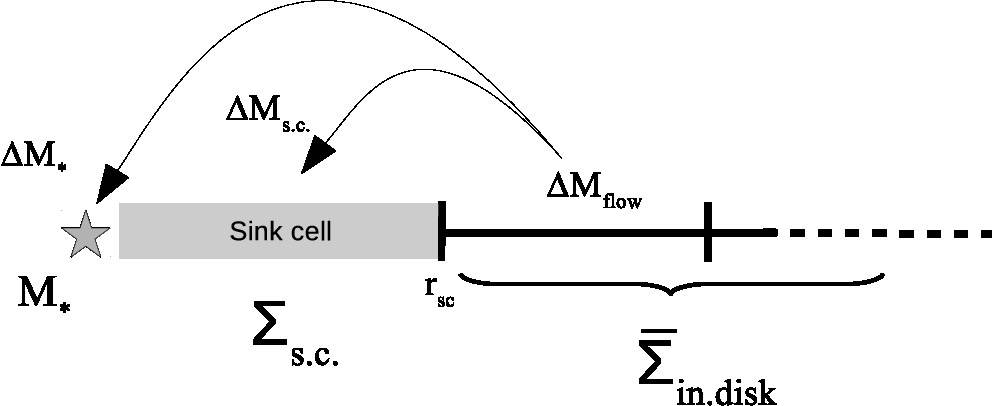}}
\par\end{centering}
\centering{}\protect\protect\protect
\caption{\label{scheme02} 
Schematic illustration of the inner boundary condition.
The mass of material $\Delta M_{\rm flow}$ that passes through the sink cell
from the active inner disk is further divided in two parts: the mass $\Delta M_\ast$ 
contributing to the  growing central star and the mass  $\Delta M_{\rm s.c.}$ settling
in the sink cell.
}
\end{figure}

We use the polar coordinates ($r,\phi$) on a two-dimensional numerical grid with
$256\times256$ grid zones. The radial grid is logarithmically spaced,
while the azimuthal grid is equispaced. To avoid too small time steps, we
introduce a ``sink cell'' at $r_{\rm sc}=1.0$~AU and impose a transparent 
inner boundary condition so that the matter (gas or dust) is allowed to
flow freely from the active domain to the sink cell and vice versa.
The mass of material that passes through the sink cell from
the active inner disk  at each time step is further redistributed
between the growing central star and the sink cell as $\Delta M_{\rm
flow}=\Delta M_\ast + \Delta M_{\rm s.c.}$ (see
Figure~\ref{scheme02}) according to the following algorithm:
\begin{eqnarray}
 \mathrm{if}\,\, \Sigma_{\rm s.c.}^n < \overline{\Sigma}_{\rm in.disk}^n\,\, \mathrm{then} \nonumber\\
 \Sigma_{\rm s.c}^{n+1}&=&\Sigma_{\rm s.c.}^n+\Delta M_{\rm s.c.}/S_{\rm s.c.} \nonumber\\
 M_\ast^{n+1}&=&M_\ast^n+\Delta M_\ast \nonumber \\
 \mathrm{if}\,\, \Sigma_{\rm s.c.}^n \ge \overline{\Sigma}_{\rm in.disk}^n\,\, \mathrm{then} \nonumber\\
 \Sigma_{\rm s.c.}^{n+1}&=& \Sigma_{\rm s.c.}^n \nonumber\\
 M_\ast^{n+1}&=& M_\ast^n + \Delta M_{\rm flow}. \nonumber
\end{eqnarray}
Here, $\Sigma_{\rm s.c.}$ is the surface density of gas/dust in the
sink cell,  $\overline\Sigma_{\rm in.disk}$ the averaged surface density
of gas/dust in the inner active disk (the averaging is usually done  over
several AU immediately adjacent to the sink cell, the exact value is
determined by numerical experiments), and $S_{\rm s.c.}$ the surface area 
of the sink cell. The exact partition between $\Delta M_\ast$ and $\Delta
M_{\rm s.c.}$ is usually set to 95\%:5\%. We note that if $\Delta M_{\rm
flow}<0$, the material flows out of the sink cell into  the active disk.
In this case, we update only the  surface density in the sink cell as
$\Sigma_{\rm s.c.}^{n+1}=\Sigma_{\rm s.c.}^n+\Delta M_{\rm flow}/S_{\rm
s.c.}$ and do not change the mass of the central star.

The calculated surface densities in the sink cell $\Sigma_{\rm
s.c.}^{n+1}$  are further used as boundary values for the surface
densities in four inner ghost zones, which have the size of the corresponding active
grid cells on the other side of the sink cell interface
and are needed for the piecewise parabolic advection scheme used in the code.  
The radial components of velocities
in the inner ghost zones are set equal to their immediate counterparts in
the active disk, while the azimuthal components in the inner ghost zones
are extrapolated assuming  Keplerian rotation as $u_{\phi,{\rm
ghost}}/u_{\phi,\rm{act}} = \sqrt{r_{\rm act}/r_{\rm ghost}} $, where
$u_{\phi,{\rm act}}$ is the $\phi$-component of velocity in the innermost
active zones, and $r_{\rm act}$ and $r_{\rm ghost}$ the radial positions
where $\phi$-components of velocities are defined in the active and ghost
zones, respectively.

These boundary conditions enable a smooth transition of surface density
between the inner active disk and the sink cell, preventing (or greatly
reducing) the formation of an artificial drop in the surface  density
near the inner boundary. We pay careful attention to ensure that our
boundary conditions conserve the total mass budget  in the system.
Finally, we note that the outer boundary condition is set to a standard 
free outflow, allowing for material to flow out of the computational
domain, but not allowing  any material to flow in.

The code was tested on the test problems applicable to the polar
geometry, showing good  performance on the relaxation and angular
momentum conservation problems \citep{VB2006}.   A small amount of
artificial viscosity added to the code to smooth shocks produces
artificial torques that are  many orders of magnitude smaller than the
physical gravitational and viscous torques \citep{VB2007}. Additional
tests pertinent to the dust component are presented in the Appendix.

\subsection{Initial conditions}
\label{initial}
Our numerical simulations start from a pre-stellar core with the radial profiles
of column density $\Sigma_{\rm g}$ and angular velocity $\Omega_{\rm g}$ 
described as follows:
\begin{eqnarray}
\Sigma_{\rm g}(r) & = & {r_0 \Sigma_{\rm 0,g} \over \sqrt{r^2+r_0^2}}\:, \\
\Omega_{\rm g}(r) & = &2 \Omega_{\rm 0,g} \left( {r_0\over r}\right)^2 \left[\sqrt{1+\left({r\over r_0}\right)^2
} -1\right],
\label{ic}
\end{eqnarray}
where $\Sigma_{\rm 0,g}$ and $\Omega_{\rm 0,g}$ are the gas surface density and angular velocity 
at the center of the core. These profiles have a small near-uniform
central region of size $r_0$ and then transition to an $r^{-1}$ profile;
they are representative of a wide class of observations and theoretical models
\citep{Andre1993,Dapp09}. The core is truncated at $r_{\rm out}=0.045$~pc, which is also the
outer boundary of the active computational domain (the inner boundary is at 
$r_{\rm s.c.}=1$~AU). The initial dust-to-gas ratio is set to 0.01 and it is assumed that
only small dust is initially present in the core ($\Sigma_{\rm d,gr}$ is 
initially set to a negligibly small value). The radial profiles of the
surface density and angular velocity of small dust are then defined as
$\Sigma_{\rm d,sm}(r)=0.01 \Sigma_{\rm g}(r)$ and $\Omega_{\rm d,sm}(r)=\Omega_{\rm g}(r)$.
The initial gas temperature is set to 20~K throughout the core.

The initial parameters $\Sigma_{\rm 0,g}=0.2$~g~cm$^{-2}$ and $r_0=1200$~AU are chosen 
so that the core 
is gravitationally unstable and begins to collapse at the onset of numerical simulations. 
The total initial mass of the core is $M_{\rm core}=1.03~M_\odot$ and
the adopted initial value of $\Omega_{\rm 0,g}=1.8$~km~s$^{-1}$~pc$^{-1}$ results in the ratio of 
rotational-to-gravitational energy  $\beta=0.24\%$, typical for pre-stellar cores \citep{Caselli2002}.

\begin{figure}
\begin{centering}
\resizebox{\hsize}{!}{\includegraphics{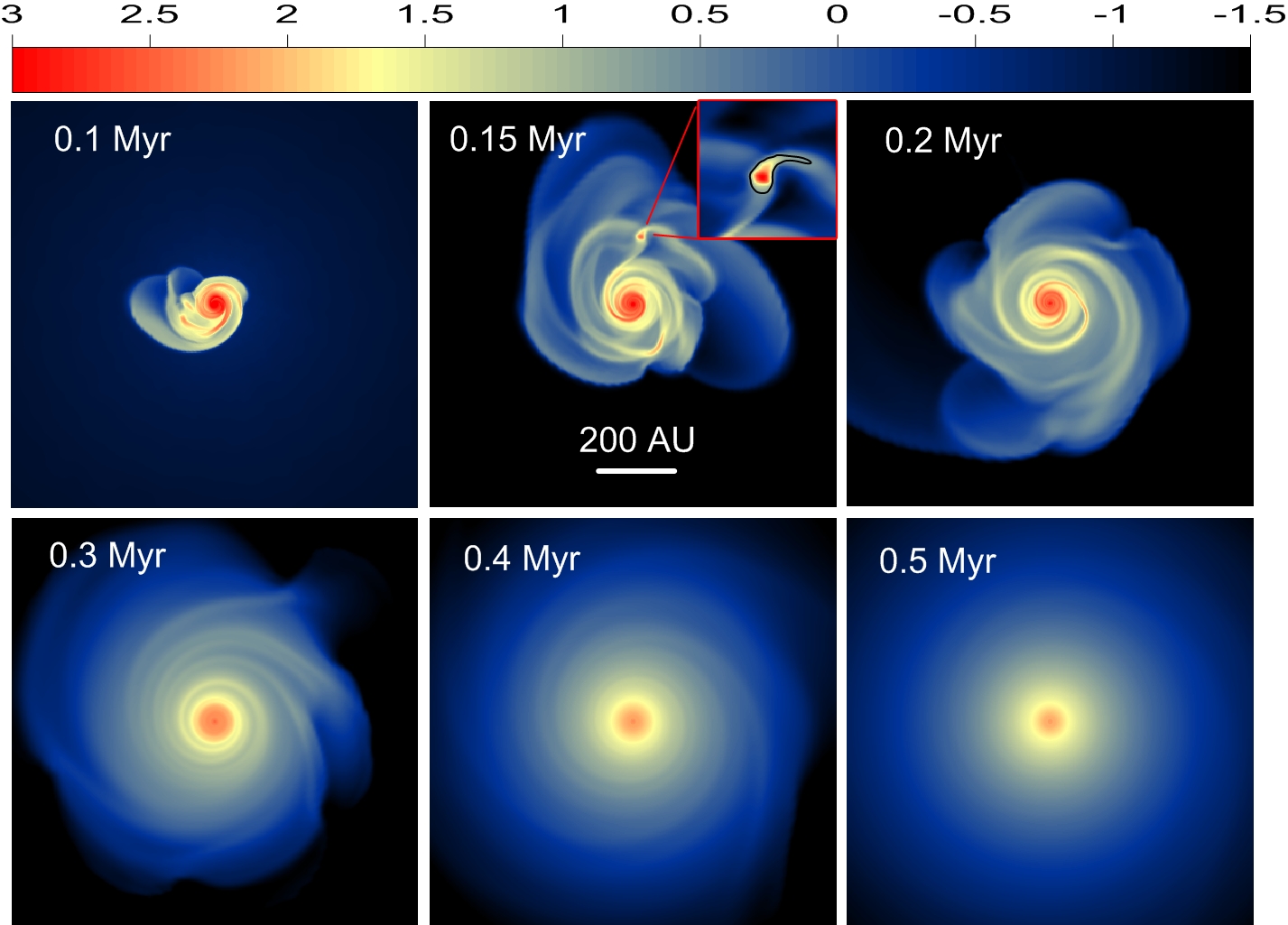}}
\par\end{centering}
\centering{}\protect\protect\protect\caption{\label{fig1} Gas surface density distribution in the inner
$1000\times1000$~AU$^2$ box at six evolutionary times starting from the onset of gravitational 
collapse. The inset shows the fragments formed in the disk via gravitational
fragmentation. The black line in the inset highlights the region where the 
Toomre parameter is less than unity. The scale bar is in log~g~cm$^{-2}$.  }
\end{figure}

\begin{figure}
\begin{centering}
\resizebox{\hsize}{!}{\includegraphics{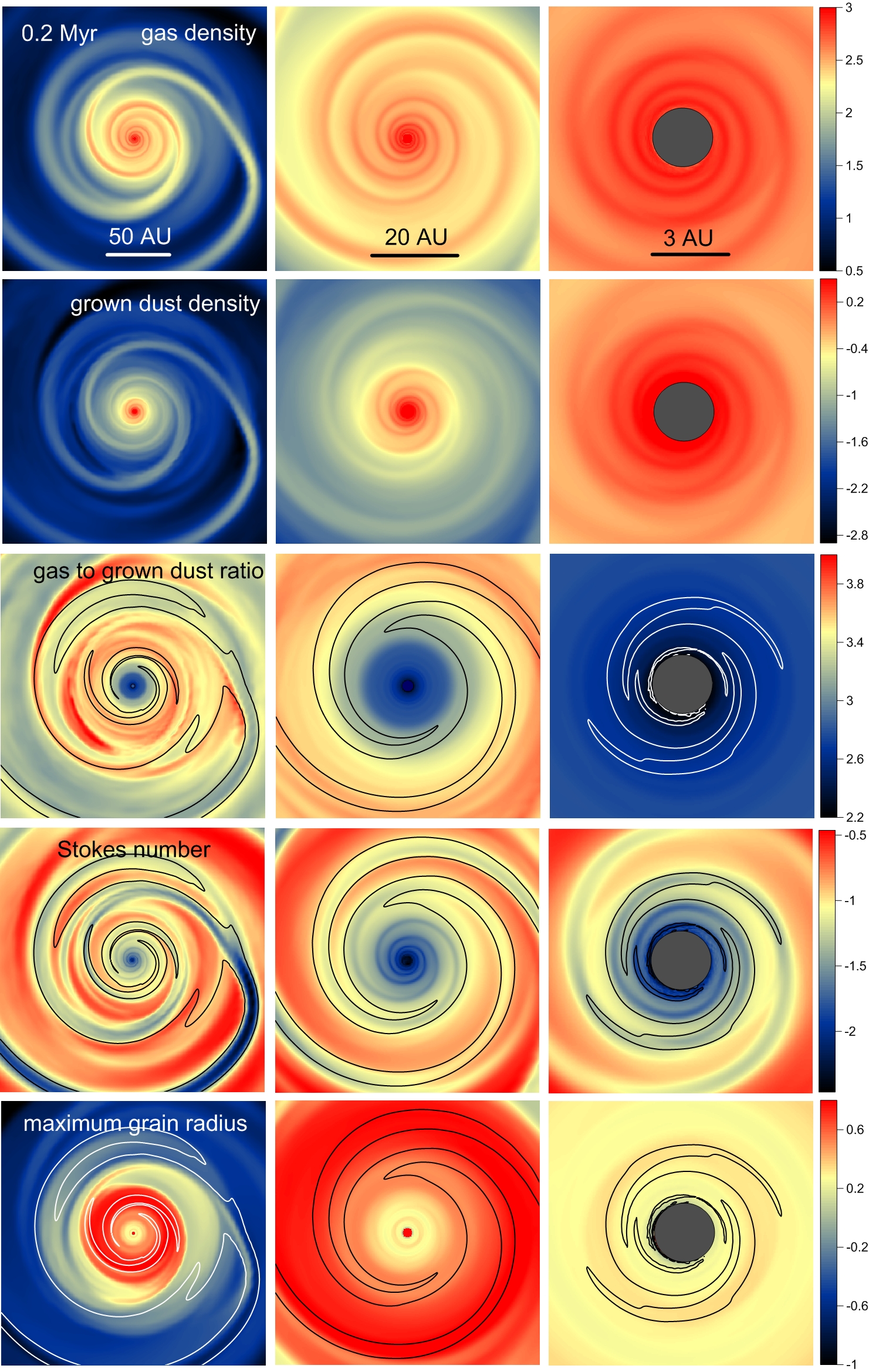}}
\par\end{centering}
\centering{}\protect\protect\protect\caption{ \label{fig2} Zoom-in onto the inner disk regions at 
$t=0.2$~Myr: left column -- $200\times200$~AU$^2$, middle column -- $60\times60$~AU$^2$, and right 
column -- $10\times10$~AU$^2$. Shown are the gas surface density in log~g~cm$^{-2}$ (top row), 
grown dust surface density in log~g~cm$^{-2}$ (second row), gas-to-grown dust mass ratio in log 
scale (third row), Stokes number in log scale (fourth row),  and maximum radius of dust grains 
$a_{\rm r}$ in log cm.  The contour lines in the third, forth, and fifth rows delineate the spiral pattern
in the gas surface density for convenience.  }
\end{figure}

\section{Main results}
\label{results}
In this section, we study the long-term evolution of a circumstellar disk formed during the gravitational
collapse of our model cloud core. 
% in the limit of a 
%constant viscous $\alpha$-parameter set to 0.01. We emphasize that we do not set a disk and star
%of a certain mass as an initial condition (as was done in many similar studies). Instead, 
%the disk and star are formed during the gravitational collapse of our model pre-stellar core. 
The disk forms at $t=0.055$~Myr and the numerical simulations are terminated at $t=0.5$~Myr. The time
is counted from  the onset of gravitational collapse ( which is also the onset of numerical 
simulations). The embedded phase of star formation
defined as the time period when the mass in the envelope is greater than 10\% of the total initial mass
in the core ends around $t=0.15$~Myr. We, therefore, cover the entire embedded phase of 
disk evolution plus some of the Class~II phase. The free parameters of our fiducial model are as follows:
$\alpha=0.01$ and $v_{\rm frag}=30$~m~s$^{-1}$. We also take into account dust settling. 
The effect of varying free parameters is considered in Sect.~\ref{paramspace}.

%\subsection{Disk evolution with constant $\alpha$-parameter}
%\label{ConstAlpha}
Figure~\ref{fig1} presents the gas surface density distribution $\Sigma_{\rm g}$ 
in the inner $1000\times1000$~AU$^2$ box. 
The computational domain covers a much larger area ($\sim 10^4\times 10^4$~AU$^2$) 
including the infalling envelope, but we show only
the most interesting inner region which covers the evolving disk. 
%The time is counted from the onset of numerical simulations. 
The central
star forms at $t=0.053$~Myr and the disk forms about 2~kyr later.  
The sequence of images illustrates how the disk grows with time and changes its shape from compact and non-axisymmetric in the embedded phase to extended and practically axisymmetric in the Class~II phase.
In the embedded phase of evolution, the disk is gravitationally unstable thanks to continuing
mass-loading from the infalling envelope, which results in the development of a notable spiral structure and even leads to episodic fragmentation. The inset at $t=0.15$~Myr zooms on to 
the gaseous clump in the $150\times150$~AU$^2$ box. 
To see if the disk fulfills the Toomre fragmentation criterion, we adopted
the modified definition of the Q-parameter appropriate for the near-Keplerian disks. 
The black contour line outlines the region where the Toomre parameter $Q_{\rm T}$ calculated 
using the following equation is less than unity:
\begin{equation}
Q_{T}={c_{\rm d} \Omega \over \pi G (\Sigma_{\rm g}+\Sigma_{\rm d,tot})},
\label{ToomreQ}
\end{equation}
where $c_{\rm d}=c_{\rm s}/\sqrt{1+\xi}$ is the modified sound speed 
in the presence of dust and
$\xi=\Sigma_{\rm d,tot}/\Sigma_{\rm g}$ the total dust to gas ratio.
Clearly, the element of the spiral arm where fragmentation took place 
is Toomre-unstable. We defer a more rigorous analysis of disk fragmentation and clump properties 
including their dust content to a follow-up, higher-resolution study. 
In the T~Tauri phase, the disk is characterized by a regular spiral structure, which slowly
diminishes as the disk loses its mass via accretion onto the star. Concurrently, the disk spreads out
due to the action of viscous torques \citep{VB2009}.

\begin{figure}
\begin{centering}
\resizebox{\hsize}{!}{\includegraphics{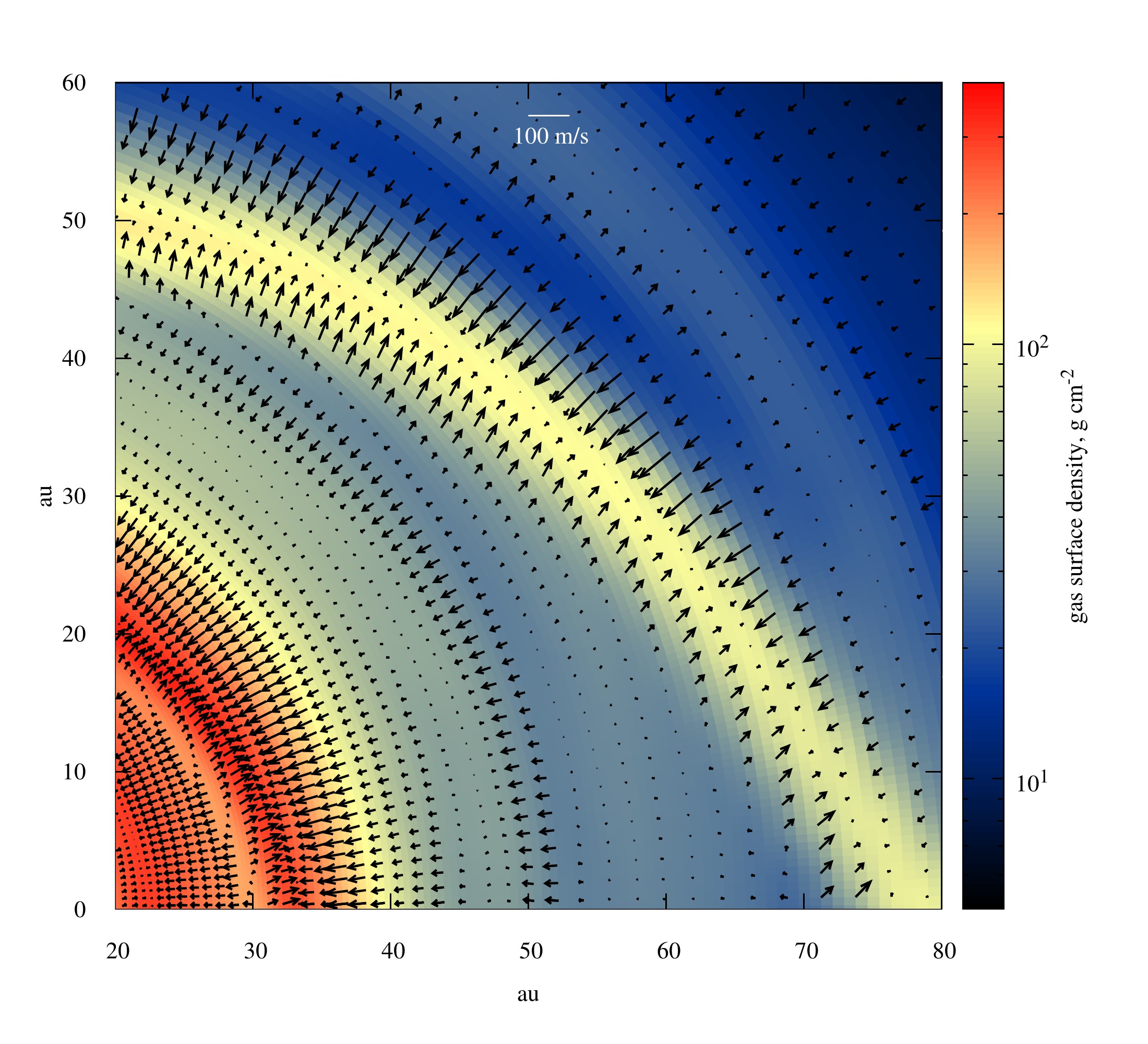}}
\par\end{centering}
\centering{}\protect\protect\protect\caption{\label{fig_vel} Velocity field for grown dust (relative to the gas) superimposed on the gas surface density map in the vicinity of a prominent spiral arm 
at $t=0.2$~Myr.}
\end{figure}

\begin{figure}
\begin{centering}
\resizebox{\hsize}{!}{\includegraphics{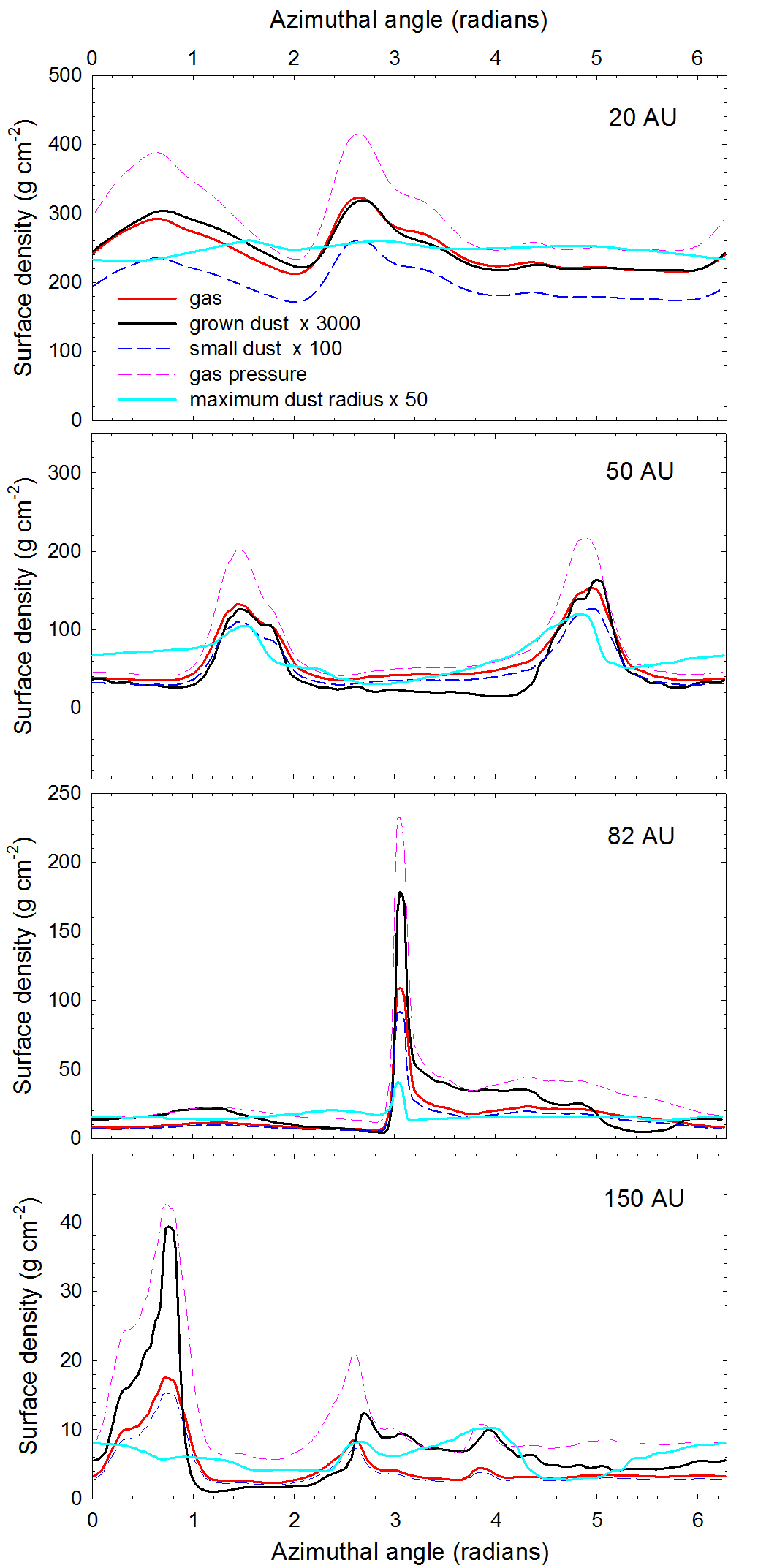}}
\par\end{centering}
\centering{}\protect\protect\protect\caption{\label{fig3b} Surface densities of gas (red solid lines), small dust (blue dashed lines) and grown dust (black solid lines) taken along the azimuthal cuts at different distance from the star: $r=20$~AU, $r=50$~AU, $r=82$~AU and $r=150$~AU. The dashed 
pink  and solid cyan lines show the gas pressure (in the code units) and the maximum radius of dust grains $a_{\rm
r}$ (in cm) taken along the same cuts. The shown evolution time is $t=0.2$~Myr.  Note that we applied
different scaling factors for the small dust density, grown dust density, and maximum dust radius.  }  
\end{figure}

In Figure~\ref{fig2} we show the zoomed-in images of the gas surface density (first row), 
surface density of grown dust (second row), gas-to-grown dust mass ratio $\Sigma_{\rm g}/\Sigma_{\rm d,gr}$ (third row), 
Stokes number (forth row), and maximum radius of grown dust (bottom row) at $t=0.2$~Myr. In particular,
the left, middle, and right columns zoom on the inner $200 \times 200$~AU$^2$, $60\times60$~AU$^2,$ and $10\times10$~AU$^2$ disk regions, respectively. The contour lines of $\Sigma_{\rm g}$ in the 
third and fourth rows of Figure~\ref{fig2} outline the position of the spiral arms, which serve as a
proxy for the position of pressure maxima (spiral arms are both denser and warmer than the inter-armed
regions). A comparison of the first and second rows indicates that the spatial distribution of 
grown dust generally correlates with the spatial distribution of gas in the sense that both exhibit a similar spatial pattern. The third row demonstrates that the inner disk 
is characterized by lower values of $\Sigma_{\rm g}/\Sigma_{\rm d,gr}$ than the outer disk, 
reflecting the process of gradual inward drift of grown dust grains. We note that
the spatial distribution of the relative content of small-to-grown dust 
$\Sigma_{\rm d,sm}/\Sigma_{\rm d,gr}$ is similar
to that of gas-to-grown dust $\Sigma_{\rm g}/\Sigma_{\rm d,gr}$ 
(but is lower by about a factor of 100) because small dust follows gas in our model.

The situation with spiral arms as the pressure maxima and the 
likely places of dust accumulation is more complicated. The spiral arms in the outer disk 
are on average characterized by lower values of $\Sigma_{\rm g}/\Sigma_{\rm d,gr}$ 
than the inter-armed regions (left panel in the third row). At the same time, 
the accumulation of
grown dust in the inner disk due to inward radial drift is more pronounced than 
accumulation of dust in the spiral arms (middle and right panels in the third row).
The fourth row of Figure~\ref{fig2} indicates that the spiral arms are characterized by
low Stokes numbers, which can be explained by higher gas densities and temperatures
in the spiral arms than in the inter-armed regions. We note that the Stokes number 
in the Epstein regime also depends linearly on the grain size, but a mild increase of $a_{\rm r}$
in the spiral arms is insufficient to compensate a stronger increase 
in gas density and temperature (see also Figure~\ref{fig3b}), so that the Stokes number 
effectively decreases.  As numerical studies of dust dynamics in 
gravitationally unstable disks \citep{Gibbons2012,Booth2016} indicate, the dust concentration 
efficiency depends on the Stokes number and is highest for the Stokes number on the order of unity.
For Stokes numbers $<0.1$,  the concentration of dust grains
in pressure maxima may be a slow process (because of rather efficient coupling with gas), 
comparable to or longer than the lifetime of the spiral pattern in our simulations.

Figure~\ref{fig_vel} presents the relative velocities between the grown dust and gas 
components superimposed on the gas surface density map. We zoom in on one of the spiral
arms in the disk (stretching from the upper-left to the bottom-right corner) to illustrate the dust drift in the vicinity of pressure maxima as represented
by the spiral arm.  The overall inward drift of grown dust towards the spiral arm is clearly seen. 
The relative drift velocities are higher at the edges of the spiral due to lower gas densities and, hence, weaker gas-to-dust coupling.

To illustrate the varying efficiency of grown dust concentration in the inner and outer spiral arms, we plot in Figure~\ref{fig3b} the surface densities of gas (red solid lines), small dust (blue dashed lines) and grown dust (black solid lines) taken along the azimuthal cuts at different distance from the star: $r=20$~AU, $r=50$~AU, $r=82$~AU and $r=150$~AU. We also plot the azimuthal profiles of
the gas pressure and maximum radius of dust grains $a_{\rm r}$. 
Spiral arms manifest themselves as local peaks in the gas surface density and pressure. The contrast
in the surface density and  pressure between the spirals and the inter-armed regions 
is smaller  and the pressure maxima are less sharp in the inner disk regions.   The azimuthal 
variations in pressure can weaken in the inner disk because spiral arms converge and wind 
up towards the disk center, as Figure~\ref{fig2} indicates.  This can also be caused by 
a weakened GI, but only in the innermost 10--15~AU where the $Q$-parameter (see
Figure~\ref{fig8}) becomes greater than 2.0. 
Clearly, the azimuthal distribution of small dust follows that of gas at all radial distances. 
On the other hand, the behaviour of grown dust is distinct from that of small dust. 
The contrast in the grown dust density between the spiral arms and the inter-armed
regions increases at larger distances and becomes higher by about a factor of two than 
the corresponding contrast in the gas surface density. At $r=82$~AU, for instance, 
$\Sigma_{\rm d,gr}$ is factors of 35 and 5 higher in the center of the spiral arm ($\phi\approx 3.0$ rad) than in the immediate inter-armed region on the left- and right-hand sides from the arm, 
respectively. The corresponding factors for $\Sigma_{\rm d,sm}$ are 18 and 6. This implies 
accumulation of grown dust in the outer parts of spiral arms.
Concerning $a_{\rm r}$, it shows little correlation with the spiral arms in 
the innermost and
outermost disk regions (20 and 150~AU, respectively). In the intermediate disk regions (50 and 82 AU), some increase in $a_{\rm r}$ at the positions of spiral arms is evident, which is
likely due to faster growth in high-density regions. However, this increase
(maximum a factor of two) is insufficient to compensate for the corresponding increase in 
the gas surface density and temperature in the spiral arms,
so that the Stokes numbers decrease in the spiral arms (see also Figure~\ref{fig2}).    

\begin{figure}
\begin{centering}
\resizebox{\hsize}{!}{\includegraphics{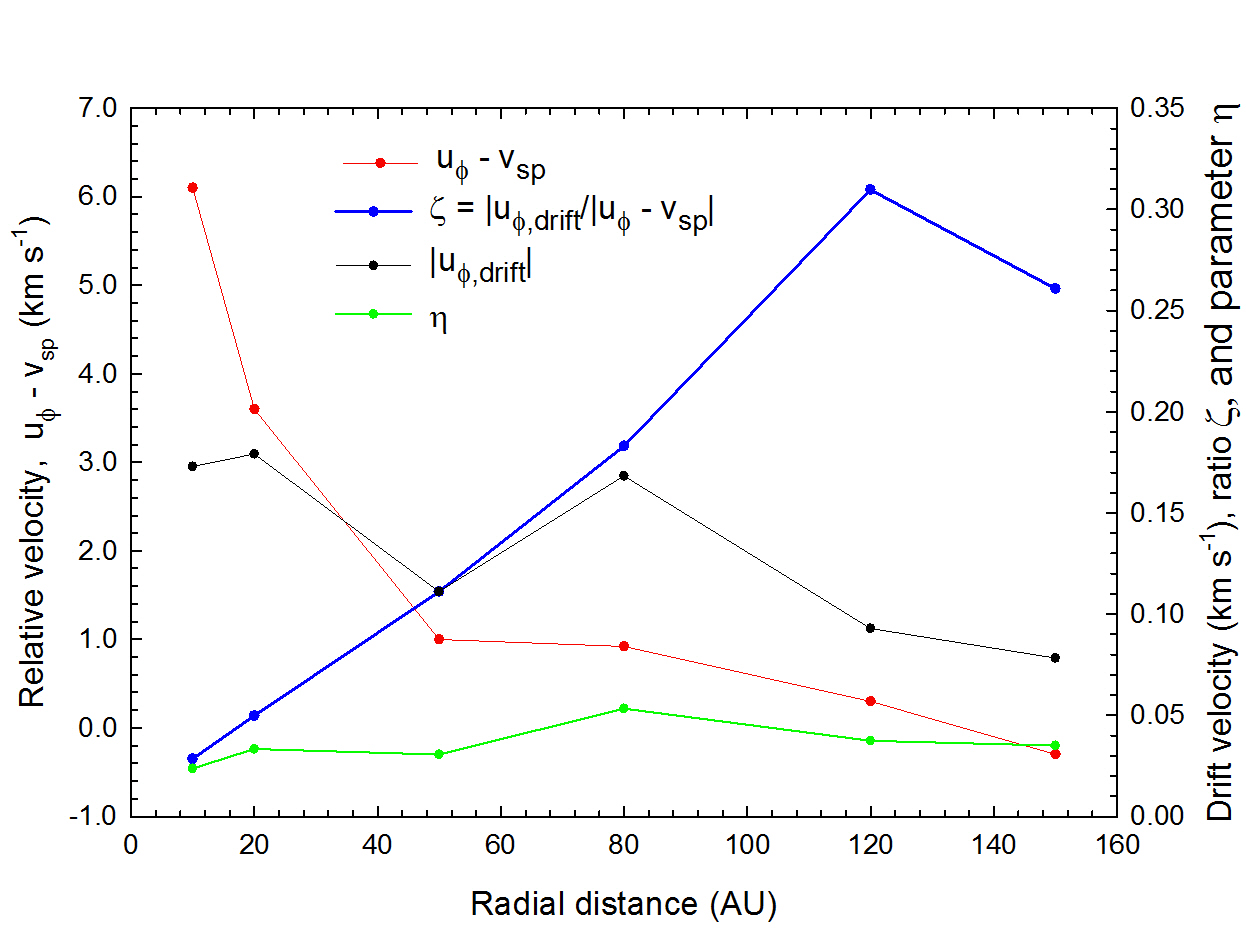}}
\par\end{centering}
\centering{}\protect\protect\protect\caption{\label{fig3c}  Radial profiles of the azimuthal component
of the dust drift velocity $|u_{\rm \phi,drift}|$,  the azimuthal component of the dust
bulk velocity relative to the local velocity of the spiral pattern  $u_\phi - v_{\rm sp}$,  
their ratio $\zeta=|u_{\rm \phi,drift}| / |u_\phi - v_{\rm sp}|$, and parameter $\eta$.  }
\end{figure}

\begin{figure}
\begin{centering}
\resizebox{\hsize}{!}{\includegraphics{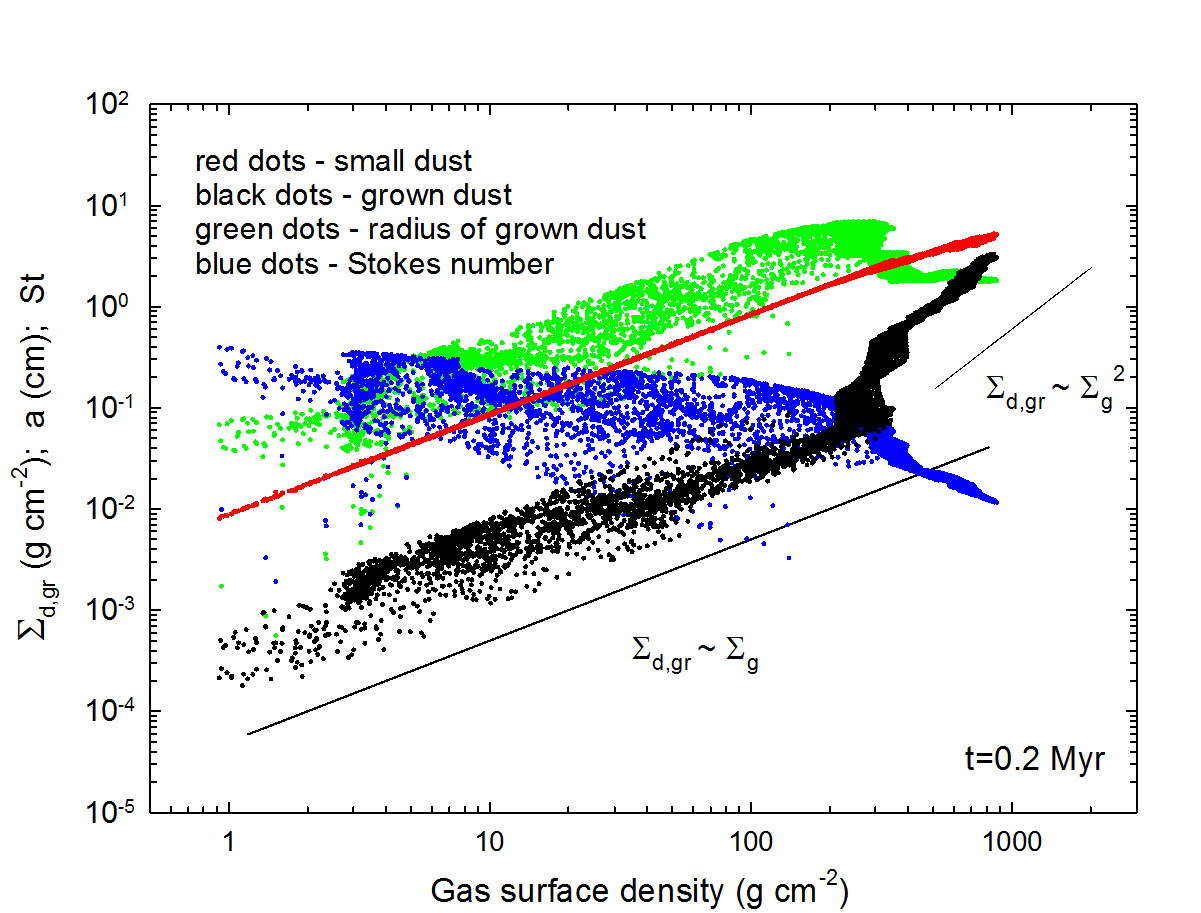}}
\par\end{centering}
\centering{}\protect\protect\protect\caption{\label{fig3} Surface density of grown dust, maximum
radius of dust grains, and Stokes number as a function of the gas surface density within 200~AU from
the star at $t=0.2$~Myr.  
The black solid lines illustrate the linear and quadratic dependencies for the ease of 
comparison.  }
\end{figure}

To understand the reason why grown dust preferentially  accumulates in the outer parts of the spiral
arms, we compare the characteristic velocities of dust drift $\bl{u}_{\rm drift}$ with the 
velocity of dust in the local frame of reference of the spiral pattern $\bl{u}_p - \bl{v}_{\rm sp}$.
The components of the drift velocity can be approximated as \citep{2016SSRv..205...41B}
\begin{equation}
u_{\rm r,drift} \simeq -{1 \over \rm{St}+\rm{St}^{-1}} \, \eta \, v_{\rm K},  \,\,\,\,  u_{\rm \phi,drift} \simeq - {1 \over 1+ \rm{St}^2} \eta\, v_{\rm K},
\end{equation}
where $v_{\rm K}$ is the Keplerian velocity and $\eta$ defines the deviation of the gas rotation velocity from the purely Keplerian law $v_\phi=v_{\rm K}\sqrt{1-\eta}$.
 In the region of interest ($r< 200$~AU), the Stokes number
varies between 0.01 and 0.2 (see Fig.~\ref{fig3}), so that the dust drift is dominated by the azimuthal component $u_{\rm
\phi,drift}$. The bulk velocity of dust is also dominated by the azimuthal component $u_\phi$ and in
the following analysis we consider only the azimuthal flow.

Figure~\ref{fig3c} presents the radial profiles of $|u_{\rm \phi,drift}|$, $u_\phi - v_{\rm sp}$, 
their ratio $\zeta=|u_{\rm \phi,drift}| / |u_\phi - v_{\rm sp}|$, and parameter $\eta$ at $t=0.2$~Myr. 
%The azimuthal component
%of the dust velocity $u_\phi$ is approximated by the Keplerian velocity and 
The velocity of the spiral
pattern $v_{\rm sp}$ is calculated by taking the azimuthal cuts at 50~yr intervals and finding the angular
velocity with which the peaks in the gas density (spiral arms) propagate at a given radial distance.
When calculating the Keplerian velocity, we also take into account the 
enclosed disk mass because our model disk is self-gravitating.
The efficiency of dust accumulation in spiral arms is maximal in the regions where 
$\zeta$ is maximal, because these are the regions where the drift velocity dominates over 
the bulk velocity of dust in the local frame of reference of the spiral pattern.
%i.e., in the regions where the difference $|u_\phi - v_{\rm sp}|$ is minimal. 
These are also the regions where the spiral pattern nearly
corotates with the azimuthal dust flow.  Note that $u_\phi - v_{\rm sp}$ is positive in the inner disk and negative in the outer disk reflecting the change of the azimuthal dust
flow near corotation -- dust moves faster than the spiral pattern inside corotation and slower outside
corotation.  The value of $\zeta$ drops at small radial distances, which explains why the 
accumulation of grown dust in the spiral arms of the inner disk regions is inefficient.
We note that $\eta$ is higher than what is usually assumed for axisymmetric 
circumstellar disks (e.g., a few per
mille, see \citet{2016SSRv..205...41B}), which is likely caused by perturbations that the spiral
arms cause to the gas flow. Finally, we note that $u_{\rm \phi,drift}$
weakly depends on the Stokes number for $\rm{St}<1.0$ and some concentration of small 
dust particles 
in the spiral arms can also be expected. This cannot be verified in our models since we
assume that small dust moves with the gas, but needs to be investigated in the future studies.

We conclude that the concentration of dust grains in the spiral arms  is most efficient near 
the corotation region, where the azimuthal velocity of dust grains is closest to the local
velocity of the spiral pattern and the ratio $\zeta$ of the dust drift velocity 
to the dust azimuthal velocity in the local frame of reference of the spiral pattern is maximal. 
The inner parts of the spiral arms are characterized by less sharp pressure maxima 
and small values of $\zeta$ so that a clear dust concentration does not have time to form.
We also note that our multi-armed spiral pattern is rather transient, as is also evident from Figure~\ref{fig1}. Dust accumulation in spiral arms should be more efficient in the presence
of a grand-design, two-armed spiral pattern as was observed in Elias~2-27 \citep{Perez2016,Tomida2017},
given that these structures live much longer than what was found in our simulations.

To further analyze the dust dynamics, we plot in Figure~\ref{fig3} the surface densities
of small and grown dust ($\Sigma_{\rm d,sm}$ and $\Sigma_{\rm d,gr}$, respectively), 
maximum radius of grown dust ($a_{\rm r}$), and Stokes number (${\rm St}$) as a function of 
gas surface density ($\Sigma_{\rm g}$)  at $t=0.2$~Myr. Only the data for the inner 200 AU around 
the star were used. The black solid lines illustrate the linear and 
quadratic functional dependence on $\Sigma_{\rm g}$. Clearly, $\Sigma_{\rm d,sm}$ 
follows the expected linear dependence on $\Sigma_{\rm g}$ 
due to strict coupling between these components. A mild weakening of $\Sigma_{\rm d,sm}$ vs.
$\Sigma_{\rm g}$ dependence in the high-gas-density limit can be explained by a 
more efficient conversion of small grains into large ones in the denser inner disk.
%grown
%dust reaching the fragmentation barrier in the inner, densest disk regions (see Figure~\ref{fig4}),
%thus increasing the local population of small dust grains.  
The surface density of grown dust $\Sigma_{\rm d,gr}$, on the other hand, reveals a more complicated pattern with the linear dependence on $\Sigma_{\rm
g}$ at the low-density tail and a quadratic or even steeper dependence in the 
high-gas-density limit. 
The likely explanation for this phenomenon is the inward radial drift, which increases the
relative abundance of grown dust in the inner disk regions characterized by the highest gas surface
densities. Moreover, the surface density of grown dust approaches that of 
small dust at the highest $\Sigma_{\rm g}$ (i.e., in the disk innermost parts). 
The Stokes numbers anti-correlate with the gas surface density,  as was already
noted in Figure~\ref{fig2}. This trend is especially pronounced in the high-gas-density limit. 
The maximum radius of dust grains $a_{\rm r}$ shows a more complicated pattern
-- it grows with $\Sigma_{\rm g}$ in  the low- and intermediate-gas-density regime, 
but saturates or even drops in the high-gas-density limit. As we demonstrate below, 
this behaviour is caused by grown dust reaching the fragmentation barrier in the inner disk.

\begin{figure}
\begin{centering}
\resizebox{\hsize}{!}{\includegraphics{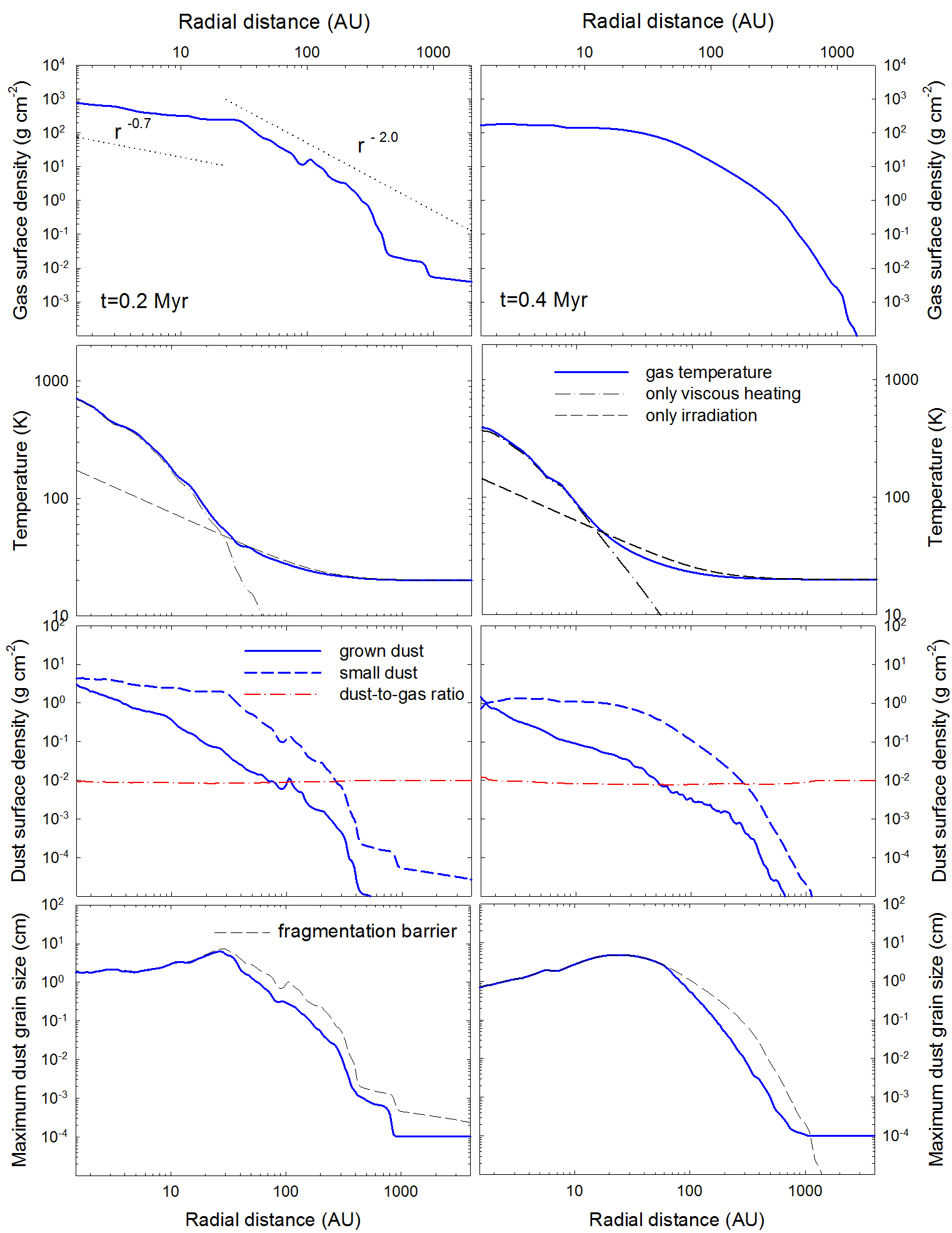}}
\par\end{centering}
\centering{}\protect\protect\protect\caption{\label{fig4} Azimuthally averaged radial profiles 
of the gas surface density (blue solid lines, top row), gas midplane temperature (blue solid lines,
second row), surface densities of small and grown dust (blue dashed and solid lines, third row), and
maximum radius of grown dust (blue lines, bottom row) at two evolutionary times: t=0.2~Myr (left
column) and t=0.4~Myr (right column). The dotted lines present various 
dependencies on radial distance $r$ for the ease of comparison. The black dashed and dash-dotted 
lines in the second row show the gas midplane temperatures as would be expected in the presence of 
either stellar+background irradiation or viscous heating only, respectively. The red dash-dotted lines
in the third row are the total dust-to-gas mass ratio. Finally, the black dashed lines in the bottom row present
the fragmentation barrier.}
\end{figure}

\begin{figure}
\begin{centering}
\resizebox{\hsize}{!}{\includegraphics{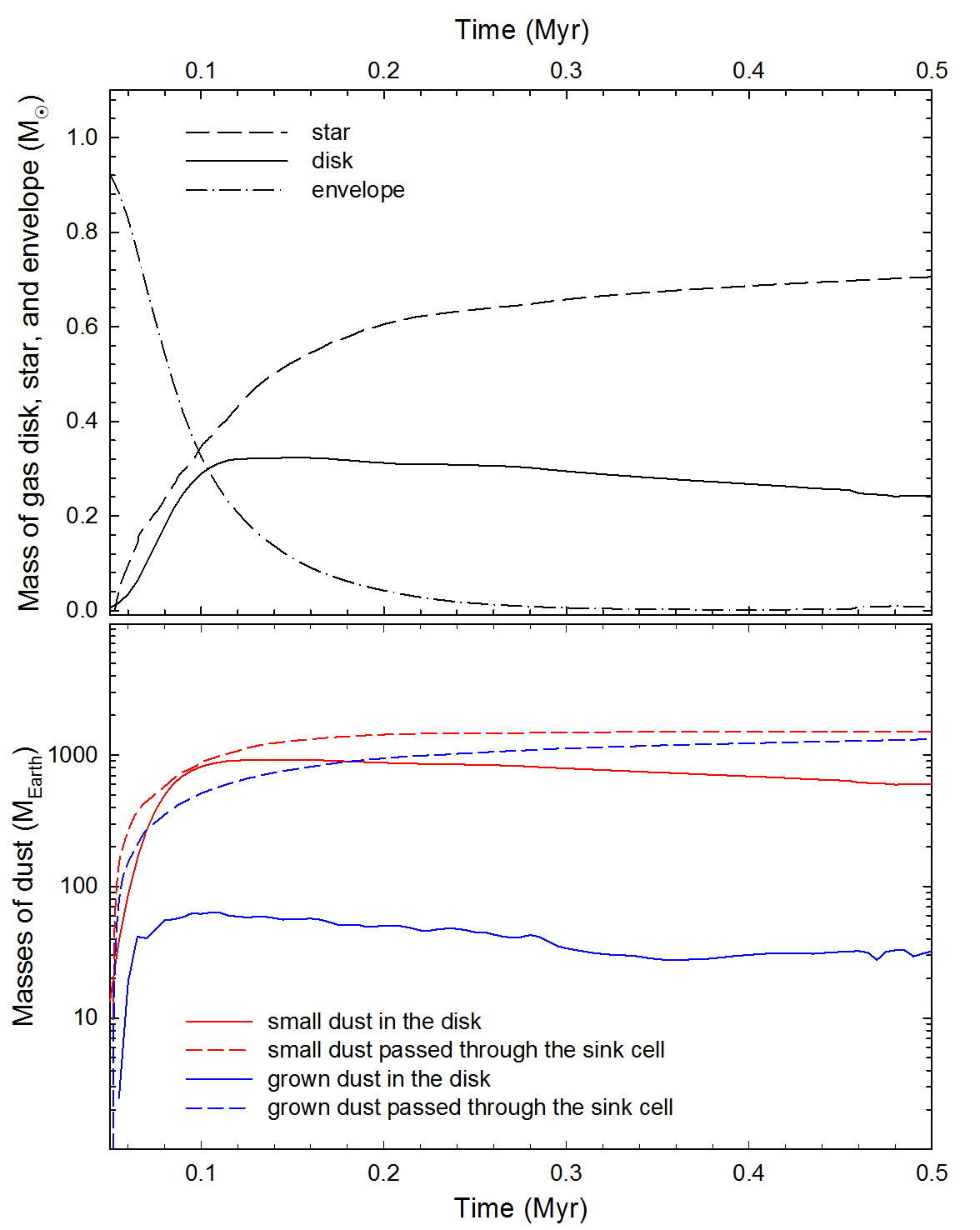}}
\par\end{centering}
\centering{}\protect\protect\protect\caption{\label{fig5} {\bf Top panel}. Integrated masses of 
the disk and envelope (solid and dash-dotted line) and the mass of the central star (dashed line) as
a function of time. {\bf Bottom panel}. Total masses of the grown dust (blue solid line) and small dust
(red solid line) in the disk as a function of time.  Also shown are the masses of small and grown
dust that passed through the sink cell (red dashed and blue dashed lines, respectively). }
%{\bf Bottom panel}. Ratio of the integrated masses of
%grown dust to small dust in the disk as a function of radial distance at $t=0.2$~Myr 
%(dashed line) and $t=0.4$~Myr (solid line).}
\end{figure}

Figure~\ref{fig4} presents the azimuthally averaged radial profiles of 
gas surface density (blue solid lines, first row), gas midplane temperature $T_{\rm mp}$ (blue solid
lines, second row), 
surface densities of small dust grains and grown dust (blue dashed and solid lines, third row), and maximum radius of dust grains (blue solid lines, bottom row). 
The black dashed and dash-dotted lines in the second row show the gas midplane temperatures
as would be expected in the presence of either stellar and background irradiation or viscous heating
only, respectively. The black dashed line in the bottom row represent the fragmentation barrier.
%The dotted lines show various power dependencies for the ease of comparison. 
The same time instances as in Figure~\ref{fig1} are shown: $t=0.2$~Myr (left column) and $t=0.4$~Myr
(right column). In the early evolution at $t=0.2$~Myr, a piecewise radial distribution 
of gas surface density is evident, with a shallower profile in the inner parts of the disk 
and a steeper profile in the outer disk. This form of $\Sigma_{\rm g}(r)$ can be expected 
for disks whose properties are governed by viscous transport 
in the inner (hotter) disk regions and by gravitational instability in the outer (colder) regions
\citep[e.g.][]{Lodato2005,VB2009}. 
%Indeed, a GI-unstable disk often features a $Q_{\rm T}$-parameter that is close
%to unity \citep[e.g.][]{Lodato2005}.  This implies that $\Sigma_{\rm g}\propto c_{\rm s} \Omega$. 
The second row demonstrates that the gas temperature in the outer disk is controlled by the stellar and background irradiation, while in the inner disk regions ($r<25$~AU)
viscous heating greatly dominates over heating due to stellar
and background radiation and viscosity becomes the dominant transport mechanism.
In the later evolution at $t=0.4$~Myr, the radial profile of $\Sigma_{\rm g}$ 
loses its piecewise 
shape. At this stage, gravitational instability subsides (as is evident in Fig.~\ref{fig1}) 
and the disk dynamics is governed mostly by viscous torques. These torques act to 
spread out the disk outer parts, leading to the gas surface density profile that is better described
by an exponent rather than by a piecewise power law \citep{LBP1974,Vorobyov2010}.

The third row in Figure~\ref{fig4} describes the azimuthally averaged characteristics of
the dusty disk component. The radial distribution of small dust $\Sigma_{\rm d,sm}$ follows that of
gas, but the radial distribution of grown dust shows notable deviations from that of gas due to inward
drift. The increase of $\Sigma_{\rm d,gr}$ towards the star is faster than that of $\Sigma_{\rm
d,sm}$, so that the ratio $\Sigma_{\rm d,gr}/\Sigma_{\rm d,sm}$ gradually rises at smaller radii. 
At a radial distance of a few~AU, $\Sigma_{\rm d,gr}$ becomes comparable to 
$\Sigma_{\rm d,sm}$.
The total dust-to-gas ratio $\xi$ shown by the red dash-dotted line, however, does not show 
significant deviations from the  initial value of 0.01.  Small grains vastly 
dominate the dust disk mass in the disk regions beyond 1.0~AU. This is a consequence of efficient inward
drift of grown dust and continuing replenishment of small dust from the infalling envelope in the embedded phase of disk evolution.

Finally, the bottom row presents the radial profile of the maximum dust radius $a_{\rm r}$. 
Clearly, the maximum growth occurs at  $r=20-40$~AU where $a_{\rm r}$ becomes as large
as several centimeters. In fact, the growth at
smaller distances is limited by the fragmentation barrier, as was also found in, e.g., 
\citet{Gonzales2015}. At larger distances $r\ga 100$~AU, 
however, dust does not grow to the fragmentation barrier and hardly exceeds 1.0~mm in radius.

Figure~\ref{fig5} presents various integrated characteristics in our model. 
In particular, the top
panel shows the masses of gas in the disk and envelope and the stellar mass as a function of
time. The bottom panel shows the masses of small and grown dust in the disk and also the mass of dust that passed through the sink cell. 
We note that the masses of gas and dust in the disk are calculated only for the disk regions beyond
$r_{\rm sc}=1.0$~AU, i.e., for the disk regions that are resolved in our numerical simulations. 
%The bottom panel 
%presents  the ratio of enclosed dust masses $M_{\rm d,gr}(r)/M_{\rm d,sm}(r)$ as a function
%of radial distance $r$ at different evolution times.
The mass of gas in the disk becomes as high as 0.3~$M_\odot$ in the embedded phase ($t\le 0.15$~Myr)
and gradually declines afterward. 
The mass of grown dust in the disk reaches 60 Earth masses in the embedded phase and drops to 30~Earth masses in the T~Tauri phase as a result of protostellar accretion.  
At the same time, the mass of small dust in the disk ($\le 1.0\mu m$) approaches 1000~Earth masses
in the embedded phase and gradually declines to 600~Earth masses at $t=0.5$~Myr,
meaning that most of the total dust mass in the disk beyond 1.0~AU actually remains in the form of small dust grains. This can also be seen from the third row of panels in Figure~\ref{fig4}.
At the same time, the mass of grown dust that passed through the sink cell is much higher than
what remains in the active disk. In fact, the masses of small and grown dust that have passed 
through the sink by the end of simulations ($t=0.5$~Myr) are comparable. 
It means that our fiducial model with $\alpha=0.01$ and $v_{\rm frag}=30$~m~s$^{-1}$ is rather efficient
in converting small to grown dust, but most of grown dust drifts quickly in the innermost, unresolved
disk regions. The fast inward drift of grown dust and continuing replenishment
of small dust from the infalling envelope (in the embedded phase) 
explains why the small dust grains vastly dominate the disk
mass beyond 1.0~AU.  We note that the process of small-to-grown dust conversion 
is very fast once the disk forms at 
$t\approx 0.056$~Myr. The mass of grown dust reaches 40~Earth masses already 
at 10~kyr after the disk formation and  a local maximum of 60~Earth masses already at 46~kyr 
after the disk formation.

\section{Parameter space study}
\label{paramspace}

\label{paramspace}
In this section, we study the effect of free parameters in our models on the distribution
of grown dust in the disk. In particular, we consider a lower value of $\alpha=0.001$ and 
lower values of the fragmentation velocity $v_{\rm frag}=10$~m~s$^{-1}$ and 
$v_{\rm frag}=20$~m~s$^{-1}$. In
addition, we consider a model without dust settling, which is motivated by the study of \citet{Rice2004}
who found that vertical stirring in a gravitationally unstable disk prevents dust particles from settling
to the midplane. This effectively means that the dust vertical scale height $H_{\rm d}$ is set equal
to the vertical scale height of gas $H_{\rm g}$.

\begin{figure}
\begin{centering}
\resizebox{\hsize}{!}{\includegraphics{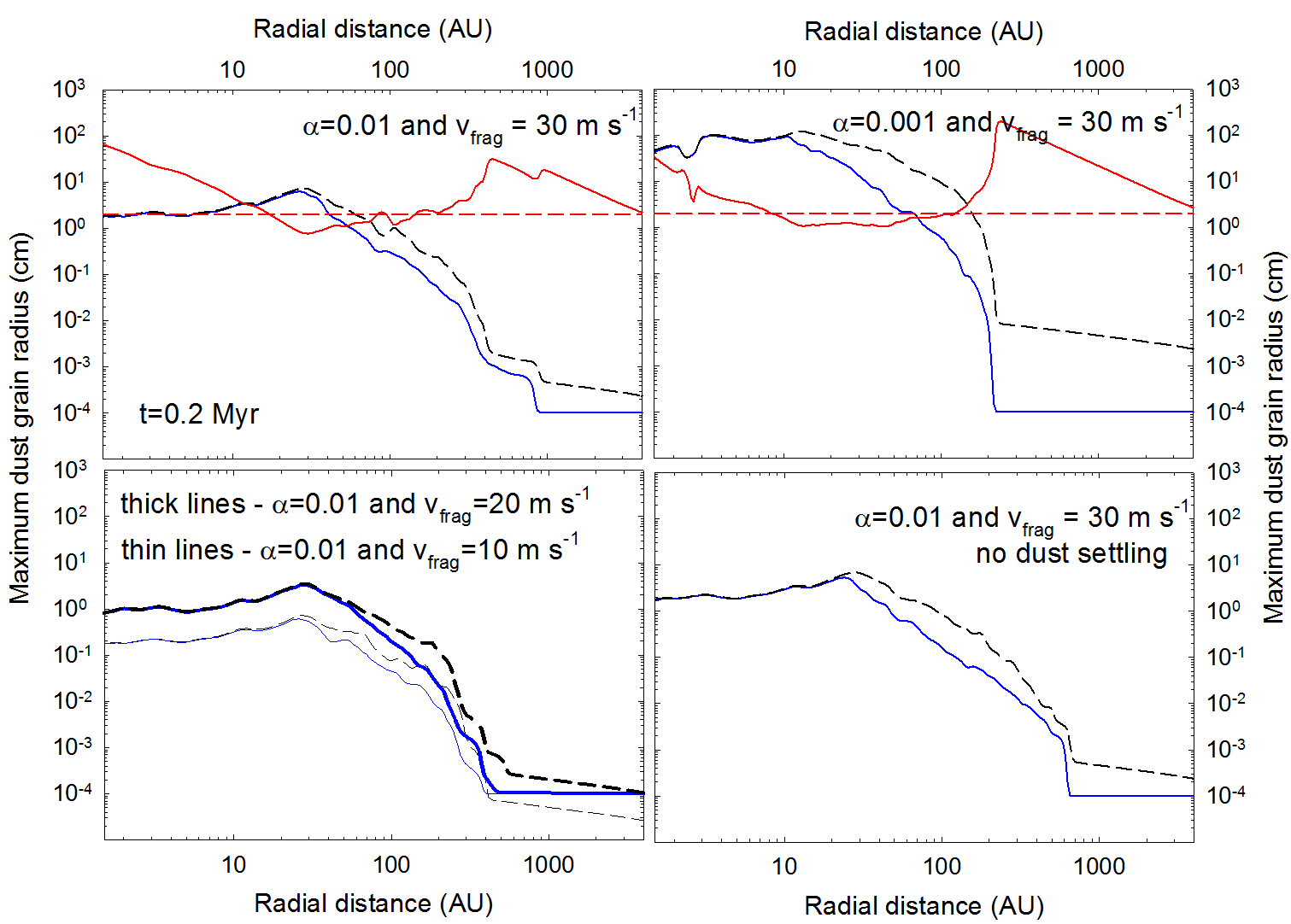}}
\par\end{centering}
\centering{}\protect\protect\protect\caption{\label{fig8} Azimuthally averaged radial profiles 
of the maximum radius $a_{\rm r}$ in models with different $\alpha$-parameters 
and fragmentation velocities as indicated in the panels. The evolution time in all panels is $t=0.2$~Myr.
%The top left panel represent the fiducial model with $
%\alpha=0.01$ and $f_{\rm frag}=30$~m~s$^{-1}$. 
The solid blue lines are the model results and the
dashed black lines are the maximum grain radius $a_{\rm frag}$ set by the fragmentation barrier
defined in Equation~(\ref{afrag}).
The red solid lines show the Toomre Q-parameter calculated using 
Equation~(\ref{ToomreQ}). The red dashed lines mark the $Q$=2.0 threshold for convenience.}
\end{figure}

Figure~\ref{fig8} presents the azimuthally averaged radial profiles of the maximum radius
$a_{\rm r}$ in models with different $\alpha$-parameters and fragmentation velocities at 
$t=0.2$~Myr. More specifically,
the top left panel corresponds to the fiducial model with $\alpha=0.01$ and $v_{\rm frag}=30$~m~s$^{-1}$.
Other panels correspond to models with $\alpha=10^{-3}$ and $v_{\rm frag}=30$~m~s$^{-1}$ 
(top right),  $\alpha=0.01$ and two values of $v_{\rm frag}=10$~m~s$^{-1}$ and $v_{\rm frag}=20$~m~s$^{-1}$
(bottom left; top curves correspond to $v_{\rm frag}=20$~m~s$^{-1}$) and to the model 
with $\alpha=0.01$ and $v_{\rm frag}=30$~m~s$^{-1}$ but without dust settling (bottom right). 
The solid blue lines are the model results and the
dashed black lines are the maximum grain radius $a_{\rm frag}$ set by the fragmentation barrier.

In the fiducial model, the dust radius does not exceed a few cm in the innermost disk, slightly increases
at  10--30~AU and gradually declines at larger radii. In the inner 30~AU, the dust growth is
clearly limited by the fragmentation barrier, whereas at larger radii the dust growth is either slow
or limited by inward drift. Turbulence affect dust growth in several ways. 
First, it facilitates dust growth via an increased collision rate. However, too large grain-grain 
relative velocities lead to destructive collisions. We do not follow individual trajectories 
of dust particles and a succession of dust growth and fragmentation events. Instead, our dust 
growth model includes a limitation on the grain size as described by Equation~(\ref{afrag}), which implicitly
takes the destructive effect of turbulence into account, albeit in a simplified manner.
Second, turbulence affects transport processes and, hence, the disk density and velocity 
structure. Third, dust grains are
efficient absorbers of free electrons~\citep{2016ApJ...833...92I} and can modulate the ionization fraction and the development of the MRI. It is therefore non-trivial to predict the exact
effect of turbulence on dust growth.

\begin{figure}
\begin{centering}
\resizebox{\hsize}{!}{\includegraphics{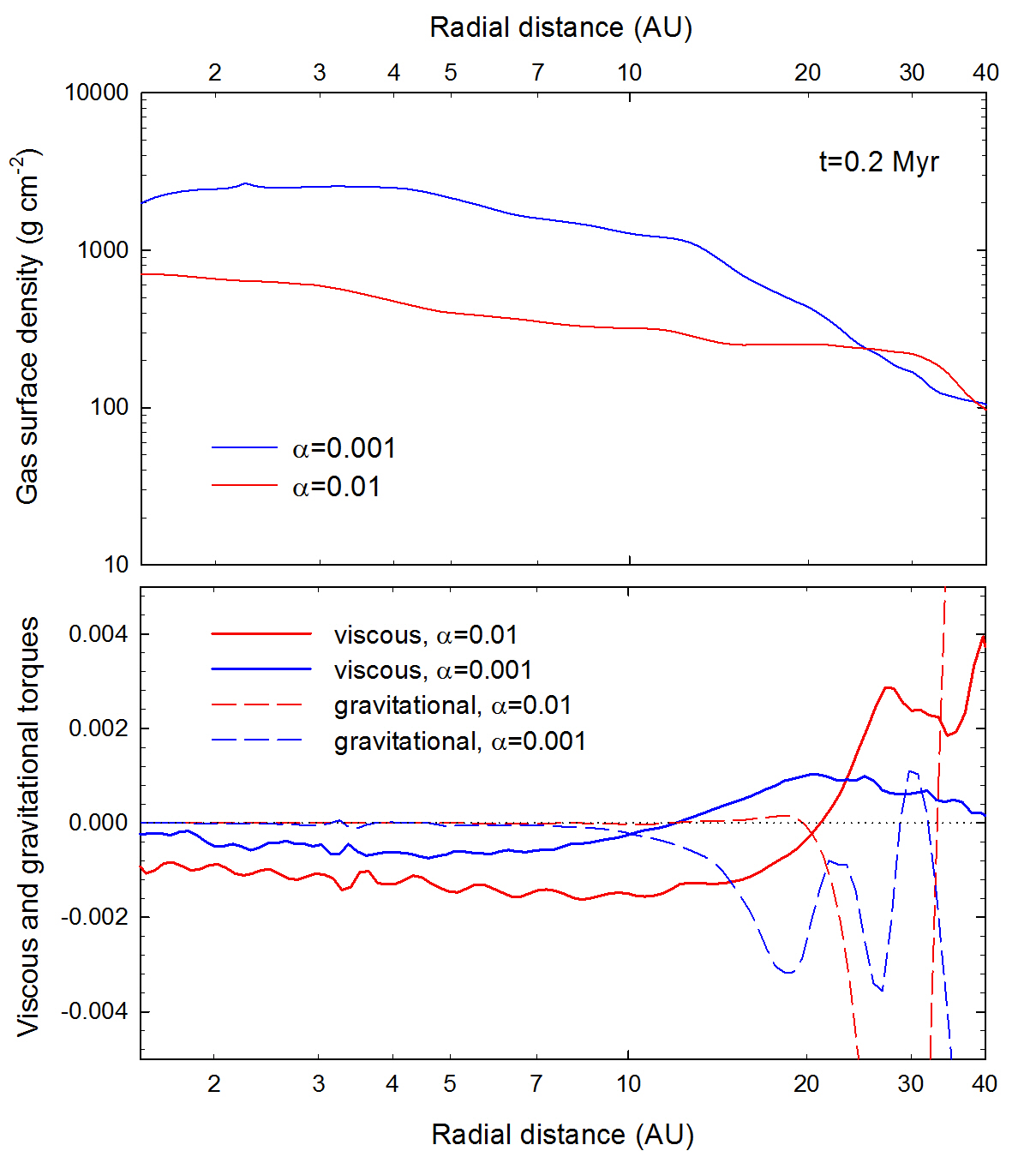}}
\par\end{centering}
\centering{}\protect\protect\protect\caption{\label{fig8a}   Azimuthally averaged profiles of 
the gas surface density, viscous torques, and gravitational torques in the inner 40~AU in the
$\alpha=10^{-3}$ model and the $\alpha=10^{-2}$ model (fiducial one) at $t=0.2$~Myr. The torques are
shown in the code units.  The black dotted lines marks the zero-torque line for convenience.}
\end{figure}

The maximum growth of dust grains is found in the $\alpha=10^{-3}$ model, 
which can be explained by a lower turbulence (and lower $\alpha$) and denser inner disk, both 
causing the fragmentation barrier to increase. 
{The gas density in the inner disk of the $\alpha=10^{-3}$ model increases due to decreased
viscous mass transport. We illustrate this phenomenon in Figure~\ref{fig8a} by comparing 
the radial profiles of the gas surface density, viscous torques, and gravitational torques 
in the $\alpha=10^{-3}$ model with the corresponding radial profiles in 
the fiducial model ($\alpha=10^{-2}$) at $t=0.2$~Myr. We focus only on the inner disk regions where
the difference in the maximum grain size is most pronounced. 
The net gravitational and viscous torques at a given radial distance $r$
are found by summing all local viscous and gravitational torques defined as
\begin{equation}
\tau_{\rm gr}(r)=\sum_\phi m(r,\phi) {\partial \Phi \over \partial \phi},
\end{equation}
\begin{equation}
\tau_{\rm visc}(r)=\sum_\phi r ({\bl \nabla} \cdot {\bl \Pi})_\phi S(r,\phi),
\end{equation}
where m($r, \phi$) is the gas mass in a cell with polar coordinates ($r, \phi$),
$\Phi$ the gravitational potential, and $S(r, \phi)$ the surface area occupied by a cell 
with polar coordinates ($r, \phi$). The summation is performed over the polar angle $\phi$ for
grid cells with the same radial distance $r$.
To reduce noise, the resulting torques were further averaged over a time period of 200~yr.
Clearly, the viscous torques in the $\alpha=10^{-3}$
model are systematically lower (by absolute value) than in the fiducial model.
 Moreover, positive values of the viscous torques are actually seen in both cases, 
but at different radii, and are to be expected when spiral arms are present. 
%Moreover, $\tau_{\rm visc}$ in the $\alpha=10^{-3}$ model becomes positive at $8<r<40$~AU, 
%indicating that viscous toques push matter outward at these distances.
At the same time,
the gravitational torques in both models are negligible (as compared to the viscous ones) 
in the inner 7~AU for the $\alpha=10^{-3}$ model and in the inner 20~AU in 
the fiducial model because of a reduced strength of GI in the inner, hot disk regions 
(see, e.g., Figure~\ref{fig2}). This is also evident from the radial profiles of the
Toomre $Q$-parameter shown by the red lines in Figure~\ref{fig8}.
%As a result, the mass transport is controlled there by viscous torques and  
%The gas density in the inner disk increases due to a juxtaposition of two effects: low $\alpha$-viscosity %and reduced strength of GI in the inner, hot disk
%regions. Indeed, the Toomre $Q$-parameter shown in the upper-right panel with the red line 
%becomes notably greater than 2.0 at $r<20$~AU, 
%implying suppression of GI in the inner disk regions.   
As a result, mass transport in the inner 10~AU of the $\alpha=10^{-3}$ model is less efficient, 
causing matter to accumulate in the inner disk and raising the gas surface density.
A similar trend of higher $\Sigma$, lower $\tau_{\rm visc}$ and negligible $\tau_{\rm gr}$ 
in the $\alpha=10^{-3}$ model holds at later evolutionary times. }
In this model, dust can grow to meter-sized boulders in the inner 10~AU. Beyond
10~AU, however, the maximum radius of dust grains drops notably below $a_{\rm frag}$. 
In the models with lower values of the fragmentation velocity shown in the bottom-left panel 
of Figure~\ref{fig8}, the dust radius is notably smaller than in the fiducial model with 
$v_{\rm frag}$=30~m~s$^{-1}$, which is clearly caused by a lower fragmentation barrier. 
We conclude that low values of $\alpha$, together with high values of $v_{\rm frag}$,  are 
needed for dust grains to grow to radii greater than a~few~cm.
Finally, we note that the presence/absence of dust settling has little effect
on the size of dust grains in the inner disk where the grain size is limited by the fragmentation barrier,
which does not depend on the dust volume density (see Equation~\ref{afrag}). In the outer disk,
the grain size in the model without settling is slightly lower than in the model with settling,
because dust growth depends linearly on the volume density of dust grains 
(see Equation~\ref{GrowthRateD}) and $\rho_{\rm d}$ is higher in the case with dust settling.

\begin{figure}
\begin{centering}
\resizebox{\hsize}{!}{\includegraphics{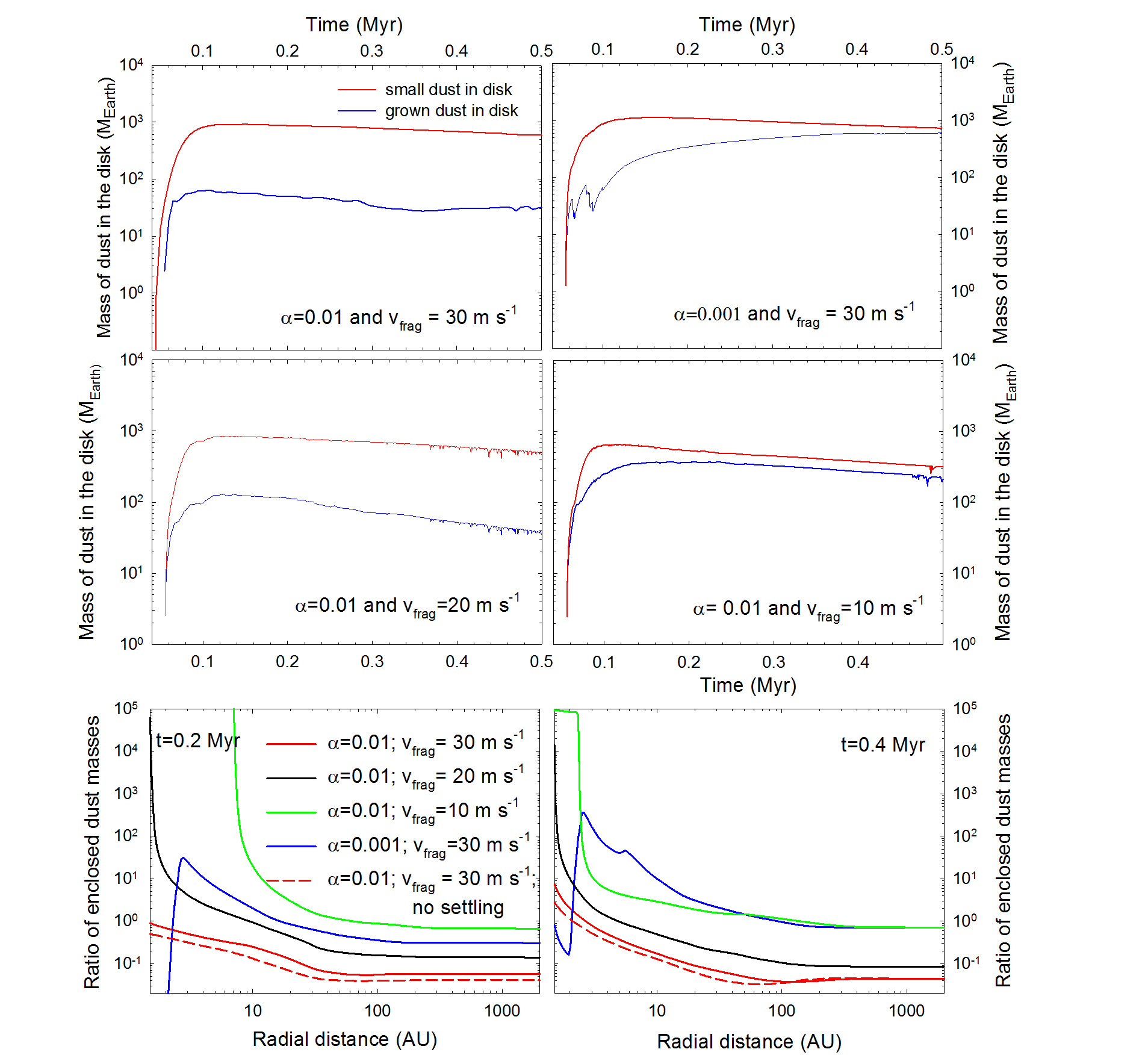}}
\par\end{centering}
\centering{}\protect\protect\protect\caption{\label{fig7} Parameter space study of dust mass in the
disk. {\bf Top and middle rows}. The mass of small and grown dust in the disk (the red and blue solid
lines) as a function of time for the fiducial model (top left) and for models with varying free
parameters as indicated in each panel. 
{\bf Bottom rows.}  The ratio of enclosed dust masses 
$M_{\rm d,gr}(r)/M_{\rm d,sm}(r)$ as a function of radial distance $r$ at two evolution times: $t=0.2$~Myr
and $t=0.4$~Myr, for the fiducial model and for models with varying free parameters.  }
\end{figure}

Figure~\ref{fig7} presents other results of our parameter space study. In particular, 
the first two rows show the total dust mass in the disk as a function of time 
for the fiducial model with $\alpha=10^{-2}$ and $v_{\rm frag}=30$~m~s$^{-1}$ (top left), for the model
with  $\alpha=10^{-3}$ and $v_{\rm frag}=30$~m~s$^{-1}$ (top right), for the model with $\alpha=10^{-2}$
and $v_{\rm frag}=20$~m~s$^{-1}$ (middle left), and for the model with $\alpha=10^{-2}$
and  $v_{\rm frag}=10$~m~s$^{-1}$ (middle right).
The bottom row presents the ratio of enclosed dust masses 
$M_{\rm d,gr}(r)/M_{\rm d,sm}(r)$ as a function of radial distance $r$ at different evolution 
times for the fiducial model and for models with varying free parameters. 
Clearly, reducing the $\alpha$-parameter has a notable effect on the total mass of grown dust in the
disk, which is higher by about an order of magnitude by the end of numerical simulations as 
compared to the fiducial case of $\alpha=0.01$. The increase is caused
by inefficient mass transport in the inner 10~AU. The resulting pile-up of 
grown dust in the inner disk (relative to small dust which traces the gas distribution) 
is clearly seen in the ratio $M_{\rm d,gr}(r)/M_{\rm d,sm}(r)$ shown by the blue lines in
the bottom row of Figure~\ref{fig7}. It is an interesting finding showing that the pile-up of 
grown dust can occur even for a constant 
but sufficiently low viscous $\alpha\le10^{-3}$ in circumstellar disks with a 
radially varying strength of GI. There is also a clear tendency for the increase in the ratio of enclosed
masses towards the star. This trend becomes more pronounced with time, reflecting gradual 
inward drift of grown dust. Our numerical
findings are in agreement with recent observations of the FU~Orionis system using VLA, which have 
found that mm- and cm-sized dust is localized in the inner several AU \citep{Liu2017}. 

It is interesting that decreasing $v_{\rm frag}$  (and making grains more prone to destruction) 
has a similar effect -- the mass of grown dust in the disk has
increased by about an order of magnitude as compared to the fiducial model. The effect is especially
pronounced in the ratio of enclosed masses $M_{\rm d,gr}(r)/M_{\rm d,sm}(r)$ shown in 
the bottom row of Figure~\ref{fig7}. In the fiducial case with $v_{\rm frag}=30$~m~s$^{-1}$ the ratio $M_{\rm d,gr}(r)/M_{\rm d,sm}(r)$ is about unity in the innermost disk and gradually declines at larger radii. In the $v_{\rm frag}=20$~m~s$^{-1}$ model, this ratio is already greater 
than unity in the inner 10~AU and increases sharply in the innermost disk regions, indicating
a more efficient conversion of small to grown dust for smaller values of $v_{\rm frag}$. 
In the $v_{\rm frag}=10$~m~s$^{-1}$ model, the inner several AU are completely
depleted of small dust grains. This counter-intuitive effect of increased small-to-grown dust conversion
for smaller values of $v_{\rm frag}$
is caused by a juxtaposition of two factors: an increased efficiency of small-to-grown dust 
conversion for lower values of $a_{\rm r}$  
and inward drift of grown dust grains. Decreasing $v_{\rm frag}$ acts to decrease
the maximum radius $a_{\rm r}$ due to a lower fragmentation barrier (see Figure~\ref{fig8}). This in turn increases the efficiency
of conversion of small to grown grains because $S(a_{\rm r})$ increases as $a_{\rm r}$ decreases 
(see Equation~(\ref{inverse2})). The grown dust drifts inward and grains
with smaller $a_{\rm r}$ from outer disk regions take their place, thus sustaining a 
cycle of dust growth. 

\begin{figure}
\begin{centering}
\resizebox{\hsize}{!}{\includegraphics{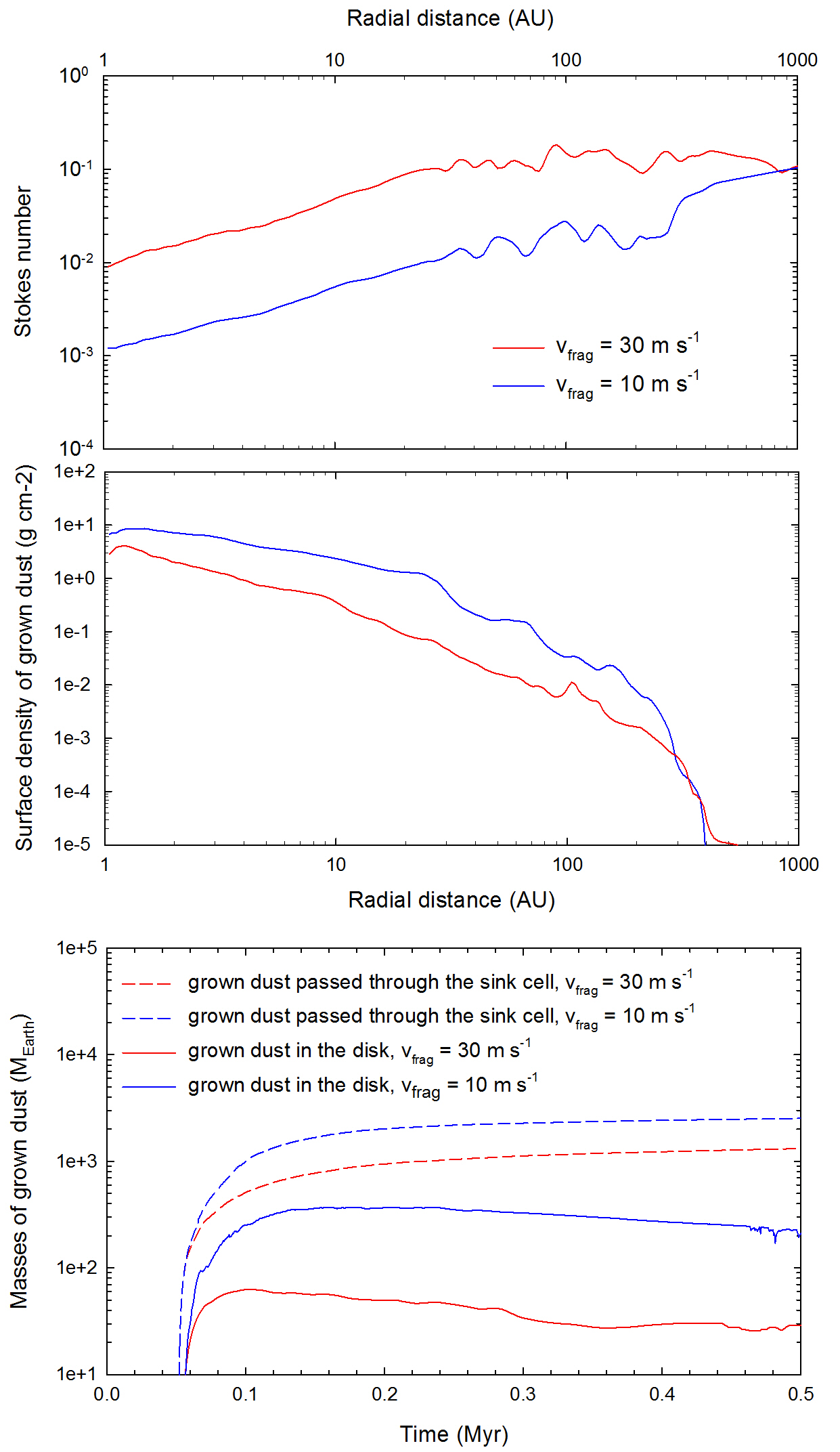}}
\par\end{centering}
\centering{}\protect\protect\protect\caption{\label{fig8b} 
Azimuthally averaged profiles of the Stokes number (top panel), surface density of grown dust 
(middle panel), and integrated masses of grown dust passed through the sink cell and 
remaining in the disk (bottom panel) in models with $v_{\rm frag}=10$~m~s$^{-1}$ (blue lines) 
and $v_{\rm frag}=30$~m~s$^{-1}$ (red lines). The radial profiles of $\Sigma_{\rm d, gr}$ and $St$ are
taken at $t=0.2$~Myr.}
\end{figure}

We further illustrate this phenomenon in Figure~\ref{fig8b} showing the azimuthally averaged profiles
of the Stokes number (top panel), surface density of grown dust (middle panel), and integrated 
masses of grown dust passed through the sink cell and remaining in the disk (bottom panel) in models
with $v_{\rm frag}=10$~m~s$^{-1}$ and $v_{\rm frag}=30$~m~s$^{-1}$.  
%reflecting a higher efficiency of small-to-grown dust conversion in this model. 
The Stokes numbers are systematically
lower in the $v_{\rm frag}=10$~m~s$^{-1}$ model because of smaller $a_{\rm r}$, indicating 
slower radial drift of grown dust.  
A less efficient dust drift in the $v_{\rm frag}=10$~m~s$^{-1}$ model can also be seen from
the ratios $M_{\rm d, gr}(\rm disk)/ M_{\rm d, gr}(total)$ of
the grown dust mass in the disk to the total produced mass of grown dust (that of residing in the 
disk and passed through the sink cell). These ratios  are 7.5\% and 2.2\% for 
the $v_{\rm frag}=10$~m~s$^{-1}$ and $v_{\rm frag}=30$~m~s$^{-1}$ models, respectively.
%indicating that the former model retains more grown dust in the disk.
This indicates that low $v_{\rm frag}$ models can be described as causing a better retention of
grown dust in the disk, as opposed to favouring a more efficient dust growth and reaching larger 
dust sizes.   
As a result, the gas surface density of grown dust increases in the $v_{\rm frag}=10$~m~s$^{-1}$
model, as can be seen in the middle panel of Figure~\ref{fig8b}.
In both models, the amount of grown dust passed through the sink
cell in the inner (unresolved) 1~AU is notably higher than what remains in the rest of the disk. 
We note that without inward dust drift this mechanism of efficient
small-to-grown dust conversion and dust accumulation would not work.
We note that fast conversion of small to grown
dust grains in the innermost disk regions was also reported by \citet{Birnstiel2010} who used 
a more sophisticated
dust growth scheme, but considered semi-analytic prescription for disk dynamics.

\begin{figure}
\begin{centering}
\resizebox{\hsize}{!}{\includegraphics{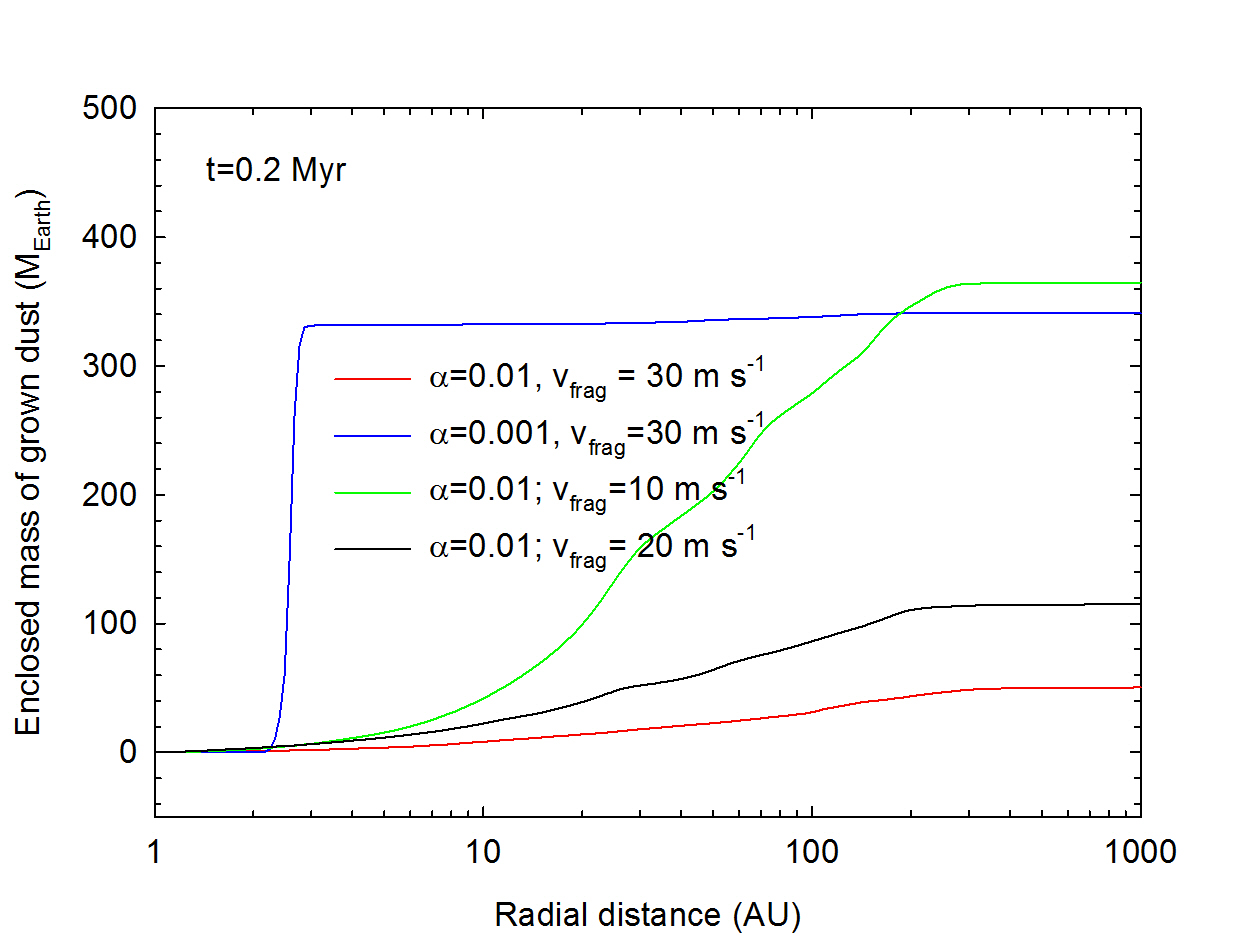}}
\par\end{centering}
\centering{}\protect\protect\protect\caption{\label{fig9} 
 Enclosed mass of grown dust as a function of radial distance at $t=0.2$~Myr for several models with
 different free parameters as indicated in the legend. }
\end{figure}

We emphasize that in all models the process of small-to-grown dust conversion is very fast. 
As Figure~\ref{fig7} indicates, the mass of grown dust reaches
tens or even hundreds of Earth masses already by the end of the embedded phase at $t=0.15$~Myr. In the model with $\alpha=10^{-3}$, the mass of grown dust continues to 
increase in the early T~Tauri stage, while in the other models it slowly decreases. The radial distribution
of grown dust is, however, distinct in different models. Figure~\ref{fig9}
shows the enclosed mass of grown dust $M_{\rm d, gr}(r)$ as a function of radial distance 
at $t=0.2$~Myr. In the $\alpha=10^{-3}$ model, almost all grown-dust mass is concentrated 
in the inner several AU, while other models
are characterized by a notably smoother radial distribution of grown-dust mass. 
It might be hard to detect grown dust in the $\alpha=10^{-3}$ model, 
due to a very limited absolute mass in grown dust grains outside 
of the inner several AU. We note that the total mass of grown dust of several hundred Earth masses 
concentrated in a narrow 
inner region with a width of several AU (as in the $\alpha=10^{-3}$ model) may be sufficient for the formation of super-Earths,
even if the conversion efficiency of dust to solid planetary cores is just a few percent.

\section{Model caveats and future improvements}
\label{caveats}

In the current model, we did not account for the momentum transfer
from dust to gas, frequently referred to as the dust ``back reaction''. As reported
in~\citet{Gonzalez2017}, this effect can be of importance
providing that conditions are met for dust trapping and subsequent growth.
We will treat this effect in the follow-up studies as soon as we
find an appropriate numerical scheme to overcome strict limitations
on the hydrodynamical time step arising when back reaction is included 
(see the Appendix).

While the adopted model of dust growth allowed us to make a first 
step towards self-consistent modeling of the gas and dust dynamics during
the long-term evolution of a protostellar disk, it is rather simplistic 
as it considers only two grain populations~-- small dust and grown dust. 
We plan to introduce more dust size bins in future studies, as was done,
e.g., in \citet{Birnstiel2010}.
The assumption of monodispersity of grown dust is hard to reconcile with
the assumption of a power-law size distribution  and it limits our treatment of grain collision 
velocities. The relative velocity is a function of size of 
colliding grains \citep{2016SSRv..205...41B} and the monodisperse approximation could 
underestimate the 
coagulation efficiency. Another critical assumption is 
the adoption of a constant (in time and space) power-law index of dust 
size distribution. This means that at every time step the dust particles are 
redistributed  over a new interval in radius ($a_{\rm min}$, $a_{\rm r}^{n+1}$) 
keeping the same power-law index $p$. This assumption could be violated as shown, 
e.g., in~\citet{Brauer2008}. In addition, other understudied 
factors such as GI-induced turbulent velocity \citep{Rice2004,Booth2016}, 
grain charging \citep{2011ApJ...731...95O,2015ARep...59..747A},
dependence of fragmentation velocity $v_{\rm frag}$ on the 
composition of dust grains, and possible temporal and spatial variations
in the viscous $\alpha$-parameter further complicate the picture. 
Such a diversity of important factors 
demands a step-by-step approach and clear understanding of 
model limitations. Therefore, we plan to refine the model 
of dust growth and complement it with a more detailed treatment of 
important processes in the near future.

\section{Conclusions}
\label{summary}

We have studied numerically the early evolution of a 
circumstellar disk formed via the gravitational collapse of a
$1.03~M_\odot$ rotating,  pre-stellar core. The evolution of the disk (beyond 1~AU) is
computed using the numerical hydrodynamics code in the thin-disk limit 
similar in methodology to the ZEUS code \citep{SN1992}, which is modified to include a dust
component. Our numerical simulations cover the embedded and early T~Tauri
phases of disk evolution and take the disk self-gravity (both of gas and
dust), friction of gas on dust, turbulent viscosity using the
$\alpha$-parameterization, and stellar irradiation into account. In this
study, we assume a spatially and temporarily constant $\alpha$-parameter.
We followed the approach of \citet{2012A&A...539A.148B} in which the
dusty disk consists of two components: sub-micron-sized dust ($\le 1.0~\mu$m)  
and grown dust ($> 1.0~\mu$m) with a maximum radius  $a_{\rm r}$ calculated using the 
method of  \citet{1997A&A...319.1007S}. Our findings can be summarized as
follows.

\begin{itemize}
\item The process of dust growth known for the older protoplanetary phase also holds 
for the embedded phase of disk evolution. In this early phase, dust growth occurs in the 
entire disk,
but its efficiency depends on the radial distance from the star -- $a_{\rm r}$ is
largest in the inner disk and gradually declines with radial distance.
In the inner 20--30~AU, $a_{\rm r}$ is limited by the assumed fragmentation barrier, while at larger
radii $a_{\rm r}$ never grows to the fragmentation barrier,
implying either slowed-down dust growth or efficient inward drift
of grown dust grains. 

\item The process of small-to-grown dust conversion is very fast once the disk is formed.
The mass of grown dust in the disk ( beyond 1~AU) reaches tens or even hundreds of Earth
masses already in the embedded phase of star formation. 
The highest concentration of grown dust is found in the $\alpha=10^{-3}$ model where 
several hundred of Earth masses can be accumulated in a narrow region of several AU from the star 
by the end of the embedded phase. Models with $\alpha=10^{-2}$ demonstrate a notably smoother
radial distribution of enclosed mass of grown dust, irrespective of the chosen 
value for fragmentation velocity. The amount of grown dust that drifts 
in the inner unresolved disk regions (inside 1~AU) by the end of our simulations (0.5~Myr)
can be even higher, on the order of thousands of Earth masses.

\item Dust does not usually grow to radii greater than a few cm. A
notable exception are models with $\alpha\la 10^{-3}$ where dust can
grow to meter-sized boulders in the inner 10 AU. 
In this case, a narrow region of several AU develops in the inner disk, which is characterized by 
reduced efficiency of mass transport via viscous and gravitational torques (low $\alpha$ and high Toomre
$Q$-parameter). This region is similar to a classic dead zone caused exclusively by varying 
viscous $\alpha$-parameter and it assists dust accumulation and growth in the 
accompanying pressure maxima.

\item The efficiency of small-to-grown dust conversion depends on the 
$\alpha$-parameter and fragmentation velocity $v_{\rm frag}$. 
In the fiducial model ($\alpha=10^{-2}$ and $v_{\rm frag}$=30~m~s$^{-1}$),  
small grains vastly dominate the dust disk mass beyond 1.0~AU. 
This is a consequence of efficient inward drift of grown dust in the inner
unresolved disk regions ($<1$~AU) and continuing replenishment of small dust from the infalling envelope.
For lower $\alpha$, grown grains dominate in the inner several AU
where a zone with reduced mass transport develops and grown dust accumulates.
For lower $v_{\rm frag}$, the conversion is also more efficient thanks to  a slower radial drift of grown dust grains due to smaller $a_{\rm r}$ and lower $St$,  
meaning a better retention of grown dust in the disk, and thanks to the inverse dependence of the dust growth rate $S(a_{\rm r})$ on $a_{\rm r}$ in our dust growth model.
For $v_{\rm frag}=10$~m~s$^{-1}$, essentially all small dust ($\le 1.0~\mu$m) is converted
to grown dust in the inner 10~AU, in agreement with \citet{Birnstiel2010}.

\item The efficiency of grown dust accumulation in spiral arms depends
on the radial position in the disk and is most efficient near corotation,
where the azimuthal velocity of dust grains is closest to the local
velocity of the spiral pattern. This is also the region where the ratio 
$\zeta$ of the dust drift velocity 
to the dust azimuthal velocity in the local frame of reference of the spiral pattern is maximal.  
For instance,  the contrast in $\Sigma_{\rm d,gr}$ between the spiral arms and the inter-armed
regions  near corotation ($r=80$~AU) is a factor of 2 higher than that 
of $\Sigma_{\rm d,sm}$, indicating grown dust accumulation.  On the contrary, the inner parts 
of the spiral arms that are located inside corotation ($r \la 50$~AU) are characterized 
by less sharp pressure maxima 
and  $\zeta$ that drops sharply, so that a clear dust concentration does not have time to form.

%Efficient inward drift of grown dust grains to the inner disk regions can also smear
%out the contrast in the density distribution of grown dust. 

\end{itemize}

 The results of our study demonstrate that the evolution of dust in early, embedded 
disks needs to be taken into account when setting the initial state for dust in older, protoplanetary disk studies. An upper limit on the dust radius of 0.25--1.0~$\mu$m usually taken for the MRN 
dust distribution \citep{MRN1997} is certainly exceeded by the end of the embedded phase.
Moreover, the maximum radius of dust grains depends on the radial position in the disk and
is not a constant in space. The radial distribution of grown dust does not follow that 
of small dust thanks to efficient inward radial drift taking place in the embedded phase.
Grown dust is more abundant in the inner than in the outer disk regions and 
an appreciable amount of grown dust drifts in the inner, unresolved 1~AU of the disk, 
probably making these disk regions more optically thick.
In the Appendix (Sect.~\ref{profiles}), we provide analytical fits to the disk state at the end of the embedded phase.

We also note that in our models the
spiral pattern in  the early embedded phase of disk evolution appears to
be volatile; its shape is stirred on timescales of a few thousand years
by differential rotation, non-linear interaction between different spiral
modes and gravitational interaction with migrating gaseous clumps. 
The process of dust accumulation in spiral arms  should be more
efficient in the presence of a grand-design, two-armed spiral pattern as
was observed in Elias~2-27 \citep{Perez2016,Tomida2017}, given that these
structures live much longer than what was found in our simulations.

%Laibe, Gonzalez \& Maddison (2012) have
%shown that there may be surface density and temperature profiles
%for which particles may survive this inward migration, and Rice
%et al. (2004, 2006), Gibbons, Rice \& Mamatsashvili (2012) have
%shown that local pressure maxima associated with density waves
%due to gravitational instabilities in the disc can trap the particles,
%saving them from the inward drift.

{\it Acknowledgements.}
We are thankful to the anonymous referee for very insightful report that helped
us to improve the manuscript and stimulated the future work on dust growth models.
This work was supported by the Russian Science Foundation grant 17-12-01168.
The simulations were performed on the Vienna Scientific Cluster (VSC-2 and VSC-3).

\begin{appendix}

\section{Analytical fits to the disk at the end of the embedded phase}
\label{profiles}
 In this section, we provide analytical fits to the disk in the fiducial model at the end of 
the embedded phase at $t=0.2$~Myr (see Fig.~\ref{fig4}). These azimuthally averaged profiles can be used as an initial 
state when modeling older, protoplanetary disks. The inner and outer disk radii are 
taken to be 1~AU and 300~AU, respectively.
\begin{eqnarray}
\overline{\Sigma}_{\rm g} \, \left[{\mathrm{g} \over \mathrm{cm}^{2}}\right]  &=&10^{2.92}\, \left({r \over \mathrm{AU}}\right)^{-0.41},  \,\,\, \mathrm{for} \,\, r < 30~\mathrm{AU}, \\
\overline{\Sigma}_{\rm g} \, \left[{\mathrm{g} \over \mathrm{cm}^{2}}\right]   &=&10^{5.65}\, \left( {r \over \mathrm{AU}} \right)^{-2.27},  \,\,\, \mathrm{for} \,\, r\ge 30~\mathrm{AU},
\end{eqnarray}

\begin{eqnarray}
\overline{\Sigma}_{\rm d,sm} \, \left[{\mathrm{g} \over \mathrm{cm}^{2}}\right]  &=&10^{0.67}\, \left({r \over \mathrm{AU}}\right)^{-0.28},  \,\,\, \mathrm{for} \,\, r < 30~\mathrm{AU}, \\
\overline{\Sigma}_{\rm s,sm} \, \left[{\mathrm{g} \over \mathrm{cm}^{2}}\right]   &=&10^{3.51}\, \left( {r \over \mathrm{AU}} \right)^{-2.23},  \,\,\, \mathrm{for} \,\, r\ge 30~\mathrm{AU},
\end{eqnarray}

\begin{equation}
\overline{\Sigma}_{\rm d,gr} \, \left[{\mathrm{g} \over \mathrm{cm}^{2}}\right]  =10^{0.9}\, \left({r \over \mathrm{AU}}\right)^{-1.56},
\,\,\, \mathrm{for} \,\, r\ge 1.0~\mathrm{AU}
\end{equation}

%\begin{eqnarray}
%\log_{10}\overline{T} \, \left[\mathrm{K}\right]  &=& 2.94 -0.38 \log_{10} \left({r \over \mathrm{AU}}\right) % ,  \,\,\, \mathrm{for} \,\, r < 30~\mathrm{AU}, \\
%\overline{\Sigma}_{\rm s,sm} \, \left[{g \over cm^{2}}\right]   &=&10^{3.51}\, \left( {r \over \mathrm{AU}} %\right)^{-2.23},  \,\,\, \mathrm{for} \,\, r\ge 30~\mathrm{AU},
%\end{eqnarray}

\begin{eqnarray}
\overline{a}_{\rm r} \, \left[\mathrm{cm}\right]  &=&10^{0.084}\, \left({r \over \mathrm{AU}}\right)^{0.41},  \,\,\, \mathrm{for} \,\, r < 30~\mathrm{AU}, \\
\overline{a}_{\rm r} \, \left[\mathrm{cm}\right]   &=&10^{4.27}\, \left( {r \over \mathrm{AU}} \right)^{-2.42},  \,\,\, \mathrm{for} \,\, r\ge 30~\mathrm{AU},
\end{eqnarray}

\begin{eqnarray}
\overline{H}_{\rm g} \, \left[\mathrm{cm}\right]  &=&10^{-0.95}\, \left({r \over \mathrm{AU}}\right)^{1.0},  \,\,\, \mathrm{for} \,\, r < 30~\mathrm{AU}, \\
\overline{H}_{\rm g} \, \left[\mathrm{cm}\right]   &=&10^{-1.8}\, \left( {r \over \mathrm{AU}} \right)^{1.46},  \,\,\, \mathrm{for} \,\, r\ge 30~\mathrm{AU},
\end{eqnarray}

\begin{eqnarray}
\overline{H}_{\rm d} \, \left[\mathrm{cm}\right]  &=&10^{-1.18}\, \left({r \over \mathrm{AU}}\right)^{0.76},  \,\,\, \mathrm{for} \,\, r < 30~\mathrm{AU}, \\
\overline{H}_{\rm d} \, \left[\mathrm{cm}\right]   &=&10^{-2.5}\, \left( {r \over \mathrm{AU}} \right)^{1.52},  \,\,\, \mathrm{for} \,\, r\ge 30~\mathrm{AU},
\end{eqnarray}

The scale heights of gas and grown dust (taking settling into account) 
are provided so that the volume distributions of gas and dust density can be recovered.

\section{Testing dust dynamics equations}
\label{Appendix}
Here, we provide the results of several essential test problems addressing the ability of
our scheme to reproduce the known analytic solutions for a mixture of gas and dust components.
While the Sod shock tube and Dusty Wave problems imply constant-size grains, 
the Dust ring problem for constant Stokes numbers allows implicitly for radial variations in the dust
size, as can be expected from radially varying disk densities, temperatures, and angular 
frequencies (see the definition of the Stokes number in Equation~(\ref{StokesN}).

\subsection{Dust ring}
For the first test problem, we consider radial drift of dust particles in a steady-state 
circumstellar disk. We make use of the analytic solution provided in, e.g., \citet{Nakagawa1986} and
\citet{Armitage2014}
%For the system (\ref{eq:system}) the quasistational approximation solution is well-known (see \cite{Nakagawa1986,ArmatageLectures2007,TakeuchiLin2002}):
\begin{equation}
\frac{u_r}{v_{K}} = - {\eta \over {\rm St} + {\rm St}^{-1}}\;.
\label{eq:analyt}
\end{equation}
%\begin{equation}
%\label{eq:vphi_analyt}
%u_{\varphi}=v_{\varphi}+\frac{u_r St}{2},
%\end{equation}
%where the small parameter $\eta$ defines the deviation of the gas rotation velocity from the 
%purely Keplerian law $v_\phi=v_{\rm K}\sqrt{1-\eta}$.
%and $r_0$ the initial location of a dust particle.
The initial radial velocity of dust  is set to zero ($u_{\rm r}=0$) and the rotation velocity is set
to the Keplerian velocity ($u_\phi=v_{\rm K}$). The mass of the central star is set to $M_\ast=0.5~M_\odot$.

\begin{figure}
\begin{centering}
\resizebox{\hsize}{!}{\includegraphics{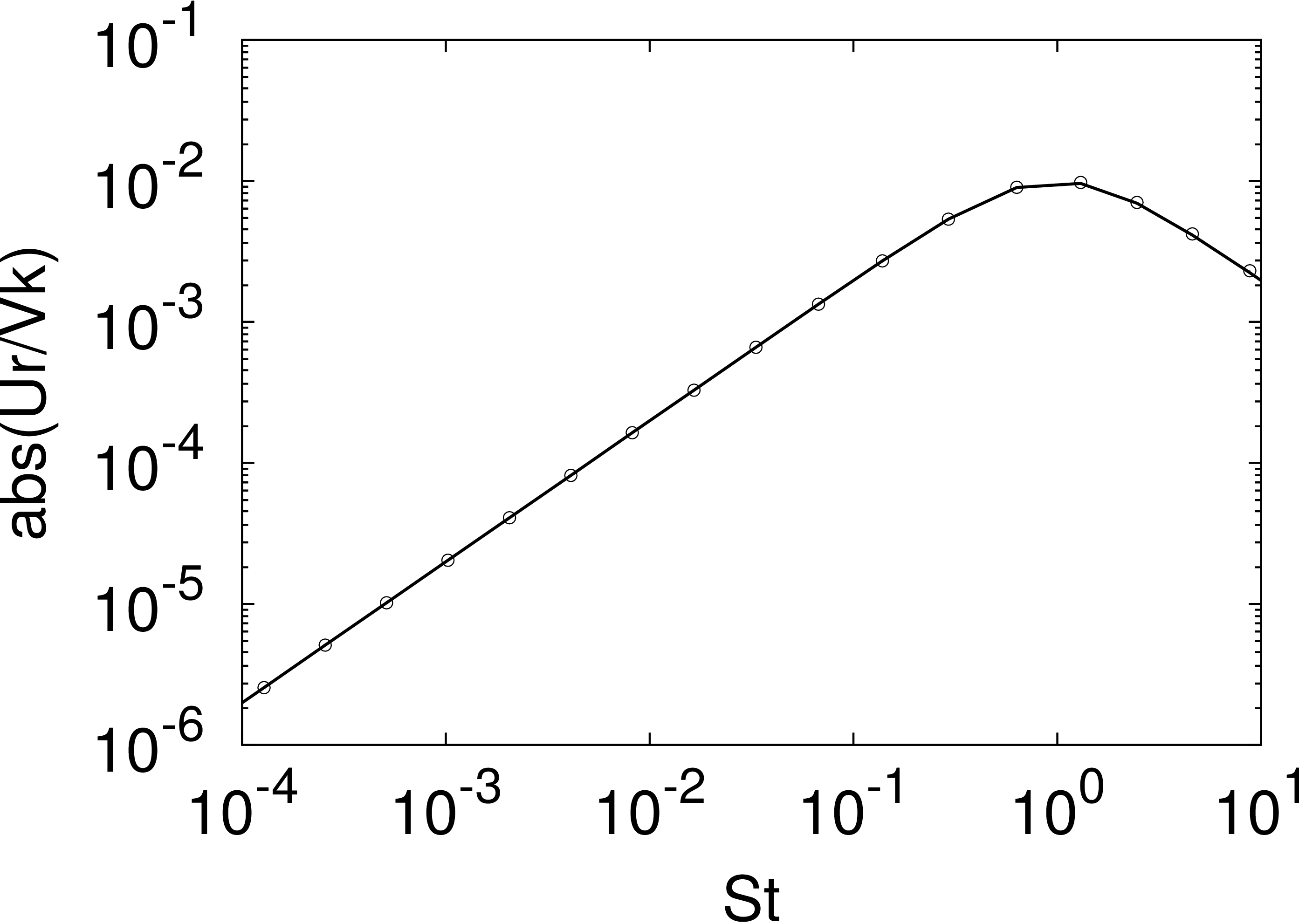}}
%\resizebox{\hsize}{!}{\includegraphics{figureA1.eps}}
\par\end{centering}
\centering{}\protect\protect\protect\caption{\label{figA1} Radial velocity of test particles in terms of Keplerian velocity. Points indicates the results of numerical simulations and the solid line 
expresses the analytical solution (\ref{eq:analyt}).}
\end{figure}

Firstly, we use test dust particles to compare the resulting drift velocity $u_{\rm r}$  
with the analytical expectations. We choose $\sqrt{1-\eta}=0.99$ and place our particles at an
initial location of $r_0=20$~AU. We allow particles to drift according the equations
\begin{equation}
\label{eq:system}
\left\{
 \begin{array}{lcl}
    
        \displaystyle 
        \frac{dr}{dt} = u_r, \\
        \displaystyle 
        \frac{du_r}{dt} =  \frac{u_{\varphi}^2}{r} - \frac{GM}{r^2} - \frac{u_r}{t_{\rm stop}},\\
        \displaystyle
        \frac{du_{\varphi}}{dt} = - \frac{u_r u_{\varphi}}{r} - \frac{u_{\varphi} - v_{\varphi}}{t_{\rm stop}}.
    \end{array}
\right.
\end{equation}
with $t_{\rm stop}$ fixed with the integration time for each particle.
The drift velocities are measured when dust particles reach a distance of 10~AU. The results for different Stokes parameters are shown in Figure~\ref{figA1}.  Clearly, our numerical scheme reproduces well the expected analytical solution for both small and large Stokes numbers.

\begin{figure}
\begin{centering}
\resizebox{\hsize}{!}{\includegraphics{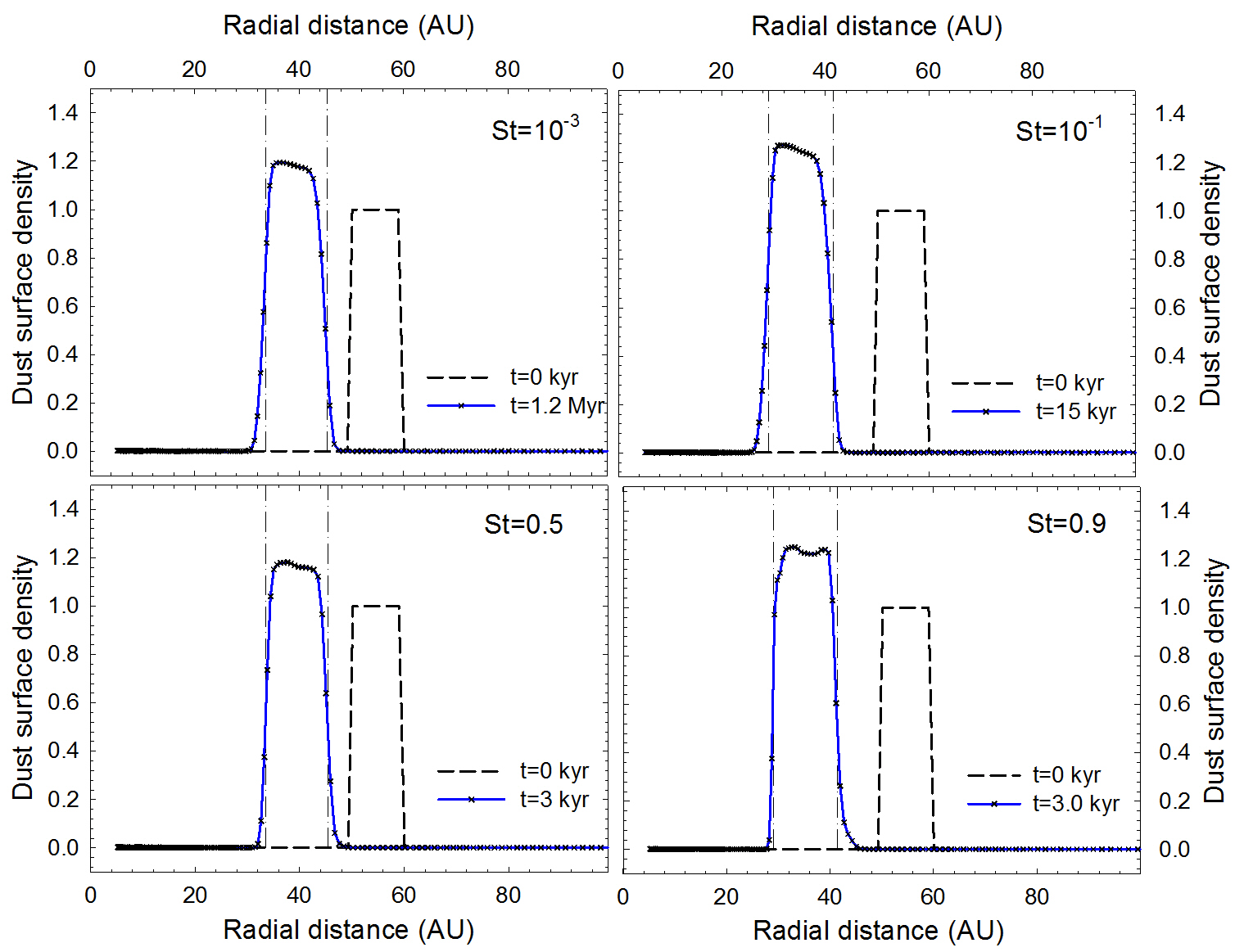}}
\par\end{centering}
\centering{}\protect\protect\protect\caption{\label{figA2} Inward drift of a dust ring. The black dashed lines show the initial position of the ring, while the blue solid lines outline the position and shape of the ring after it drifted inward by more than its full width. The vertical dash-dotted lines present the analytic solutions for the edges of the ring.}
\end{figure}

In the second step, we test the ability of our numerical hydrodynamics code to transport a dust ring across the disk.
We set a dust ring with a unit surface density and a width of 10~AU located initially between 50~AU and 60~AU. We set $\sqrt{1-\eta}=0.99$ and 
compute the dynamics of the dust ring for different values of the Stokes number. We keep the Stokes parameter fixed in space and time during the integration. The results are shown in Figure~\ref{figA2} for runs with ${\rm St}=10^{-3}$, ${\rm St}=10^{-1}$, ${\rm St}=0.5$, and ${\rm St}=0.9$. The black dashed
lines outline the initial shape and position of the dust ring and the solid blue lines outline
the ring after it drifted toward the star by more than its full width. The vertical dash-dotted
lines present the analytic solutions for the edges of the ring \citep[see, e.g.,][]{SSV2017}: 
\begin{equation}
\label{eq:rad_analyt_gen}
r^{3/2}=r_0^{3/2}-\frac{3 \sqrt[]{GM_\ast} \eta t}{2 ({\rm St}+{\rm St}^{-1})}.
\end{equation}
Evidently, the position of the ring
is reproduced rather well for all Stokes numbers that are pertinent to our modeling. The edges of the ring are smeared over several grid zones, which is consistent with the accuracy of our advection scheme. We note that the peak value of the ring increases as it drifts inward because of the shrinking surface
area of the ring.

\subsection{Sod shock tube}
\label{sec:ShockTube}

This test is often used to assess the ability of a numerical algorithm
to accurately track the position of moderate shock
waves and contact discontinuities. Initial conditions involve two
discontinuous states along the z-axis, with a hot dense gas on
the left and cold rarified gas on the right. More specifically,
we set the pressure and density of gas at $z\in[0.0.5]$ to 1.0, while at
$z\in[0.5.1.0]$ the gas pressure is 0.1 and gas density is 0.125. The velocity
of a $\gamma$=1.4 gas is initially zero everywhere. The dust component has the
same initial distribution as that of gas, but the dust pressure is set to zero.
The dust-to-gas ratio is, therefore, equal to unity everywhere.

\begin{figure}
\begin{centering}
\resizebox{\hsize}{!}{\includegraphics{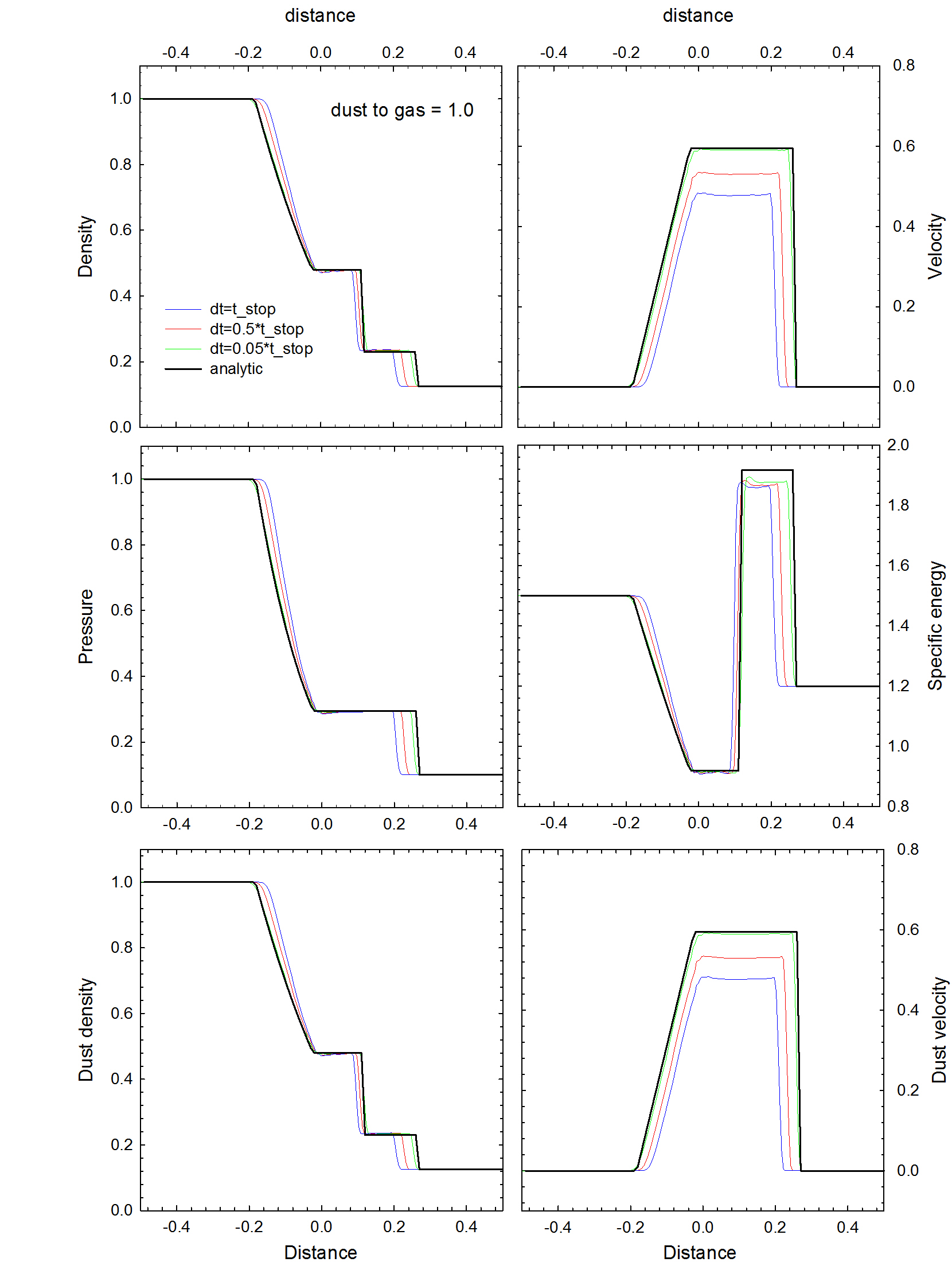}}
\par\end{centering}
\centering{}\protect\protect\protect\caption{\label{figA3} Sod shock tube problem in the limit of a
short
stopping time and taking the back reaction of dust on gas into account. The top and middle 
panels present solutions for the gas component, while the bottom rows show solutions for the dust component.
The black solid line shows the analytic solution and the
color lines present numerical solutions for different values of the time step as indicated in the top
left panel.
 }
\end{figure}

The analytic solution for the gas and dust mixture is only known in the limit of short stopping times compared to the time of shock wave propagation. We use the SPLASH code \citep{Price2007} to generate the analytic solution (on its stationary stage) with a modified sound speed for the gas-dust medium $c_{\rm d}$ (see Sect.~\ref{results}). In this case, the dust velocity is believed to be equal to the gas velocity, when the density of the dust is equal to the the density of the gas multiplied by initial dust-to-gas ratio. We follow the terminology of \citet{Laibe2012} and introduce the drag
parameter $K$ and define the stopping time as $t_{\rm stop}=\rho_{\rm gas}/K$, where $\rho_{\rm gas}=1$ is the gas density in left half of the shock tube. For this test problem, we also included the back reaction of dust onto gas by adding the corresponding term to the gas dynamics equation. Figure~\ref{figA3} present the test results for $K=1000$. The first and second rows present the resulting gas distributions at t=0.24, while the bottom row shows those of dust. The numerical resolution is 200 grid zones and
the artificial viscosity parameter is set to $C2=2$, implying that shocks are spread over two grid zones. 

We have found that in order to achieve a reasonable agreement with the analytic solution, one needs a time step $dt$ that is much smaller than the stopping time. For instance, the blue lines present the test results for $dt=t_{\rm stop}$, which obviously show considerable deviations from the analytic solution, but the green lines corresponding to $dt=0.05 t_{\rm stop}$ match the analytic solution much better (still some deviations, particularly, in the specific energy are notable). Similar results were obtained for $K=100$. 

This problem with the numerical scheme is analogous to that highlighted in \citet{Laibe2012}, but they formulated it in terms of spatial resolution, while we think it is better to express it in terms of the time step and stopping time. Anyway, the time step is directly linked to the numerical resolution for explicit numerical hydrodynamics solvers, such as our own, so the two approaches are similar. 

Our test essentially demonstrates that for the medium with high concentration of dust particles strongly coupled to the gas, the stopping time has to be resolved by at least 50 time steps to adequately track shock waves that can be present in GI unstable disks.

\begin{figure}
\begin{centering}
\resizebox{\hsize}{!}{\includegraphics{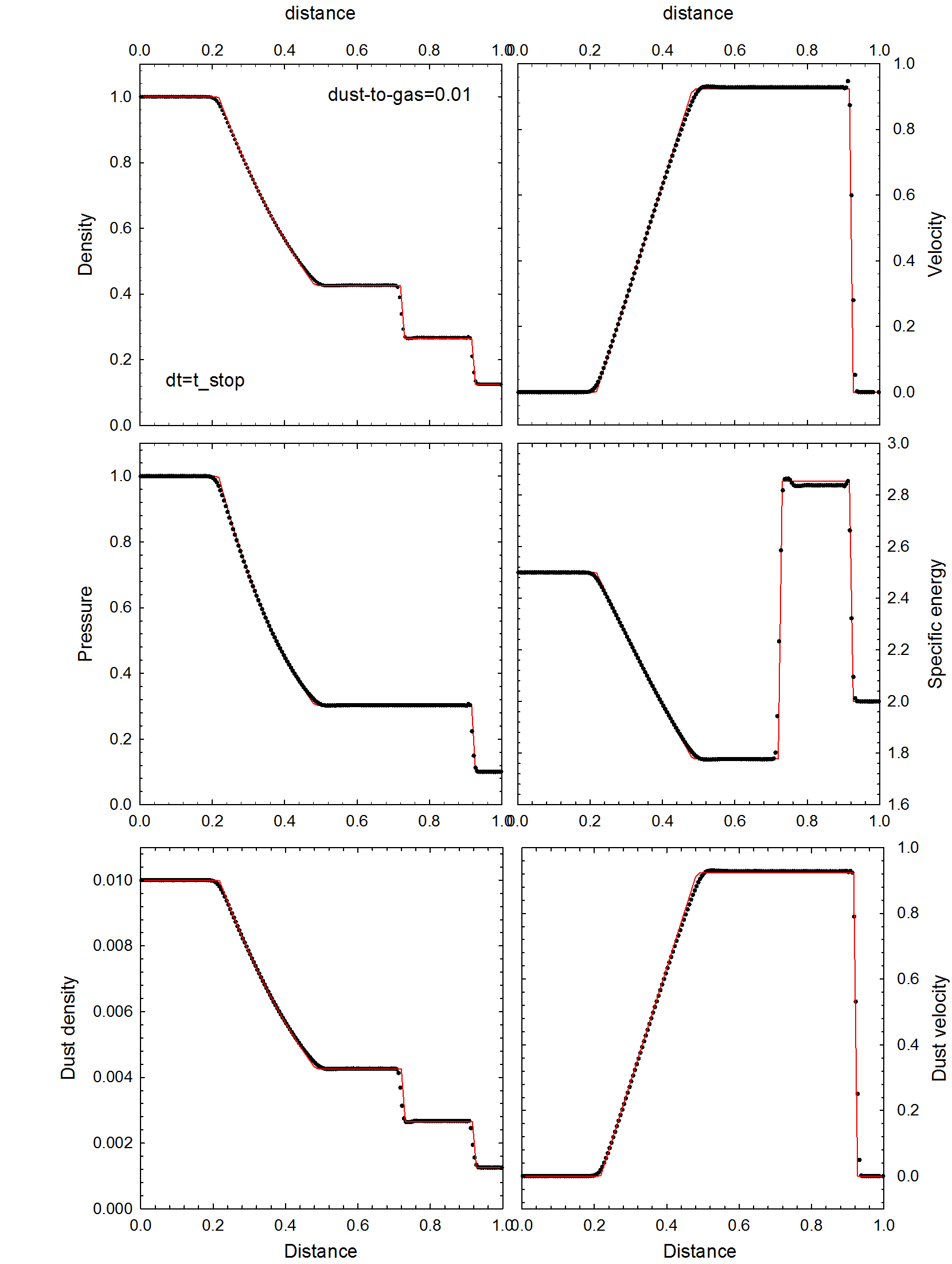}}
\par\end{centering}
\centering{}\protect\protect\protect\caption{\label{figA4} Sod shock tube problem in the limit of a short stopping time, but without taking the back reaction of dust on gas into account. The top and middle panels present solutions for the gas component, while the bottom rows show solutions for the dust component.
The red lines present the analytical solution and the filled circles are the numerical solution for the time step similar to the stopping time.
}
\end{figure}

Fortunately, we have found that our scheme works much better if the back reaction of the dust component on the gas is neglected (as we did in the present study).
Figure~\ref{figA4} presents the corresponding results for the dust to gas ratio of 0.01 when the reaction of dust on the gas component has a small effect and can be neglected. The red lines provide the analytic solution at $t=0.24$, 
while the filled circles are the numerical solution. The time step in this case is similar to the stopping time. Evidently, the code reproduces quite well the analytic solution in both gas and dust. Similar results were obtained  for $dt=2 t_{\rm stop}$ by decreasing the numerical resolution by a factor of 2. We cannot test the scheme for even larger time steps because of the Courant limitation on the time step, but this test already demonstrates that if the back reaction of dust on gas can be neglected, then there is no need to introduce another, very stiff limitation on the time step to resolve the stopping time by at least 50 time steps.

\subsection{Dusty wave}

The dusty wave problem tests the propagation of linear waves in the mixture of gas and dust. 
Unlike the Sod shock tube problem, where the analytical solution for the weak coupling between gas and
dust (low drag parameter $K$) is not known, the dusty wave problem has analytical solutions 
for both weak and strong coupling. 
%It makes possible to verify the results of simulation of dust and gas mixtures both on transient and %stationary stage. 
In particular, this test allows us to estimate the accuracy with which gas densities, gas and dust velocities are calculated during the propagating and damping of acoustic waves.

The initial setup consists of a sinusoidal wave  of the following
form
\begin{equation}
\label{eq:DustyWave_init1}
\rho_0=\tilde{\rho_{\rm g}}+\delta \sin(kx), \ \rho_{\rm d,0}=\tilde{\rho_{\rm d}}+\delta \sin(kx),
\end{equation}
\begin{equation}
\label{eq:DustyWave_init2}
v_0=\delta \sin(kx), \ u_0=\delta \sin(kx).
\end{equation}
We adopt isothermal gas  with $c_s$=1, $\tilde{\rho_{\rm g}}=1$, $k=1$, and $\delta=0.01$ 
throughout this section and vary $K$ and $\tilde{\rho_{\rm d}}$. All simulations were done with 
100 grid zones on unit distance with artificial viscosity parameter $C2=2$ and periodic boundary 
conditions. The analytical solution  is given in \citep{Laibe2011}; we used the code accompanying 
their paper.

\begin{figure}
\begin{centering}
\resizebox{\hsize}{!}{\includegraphics{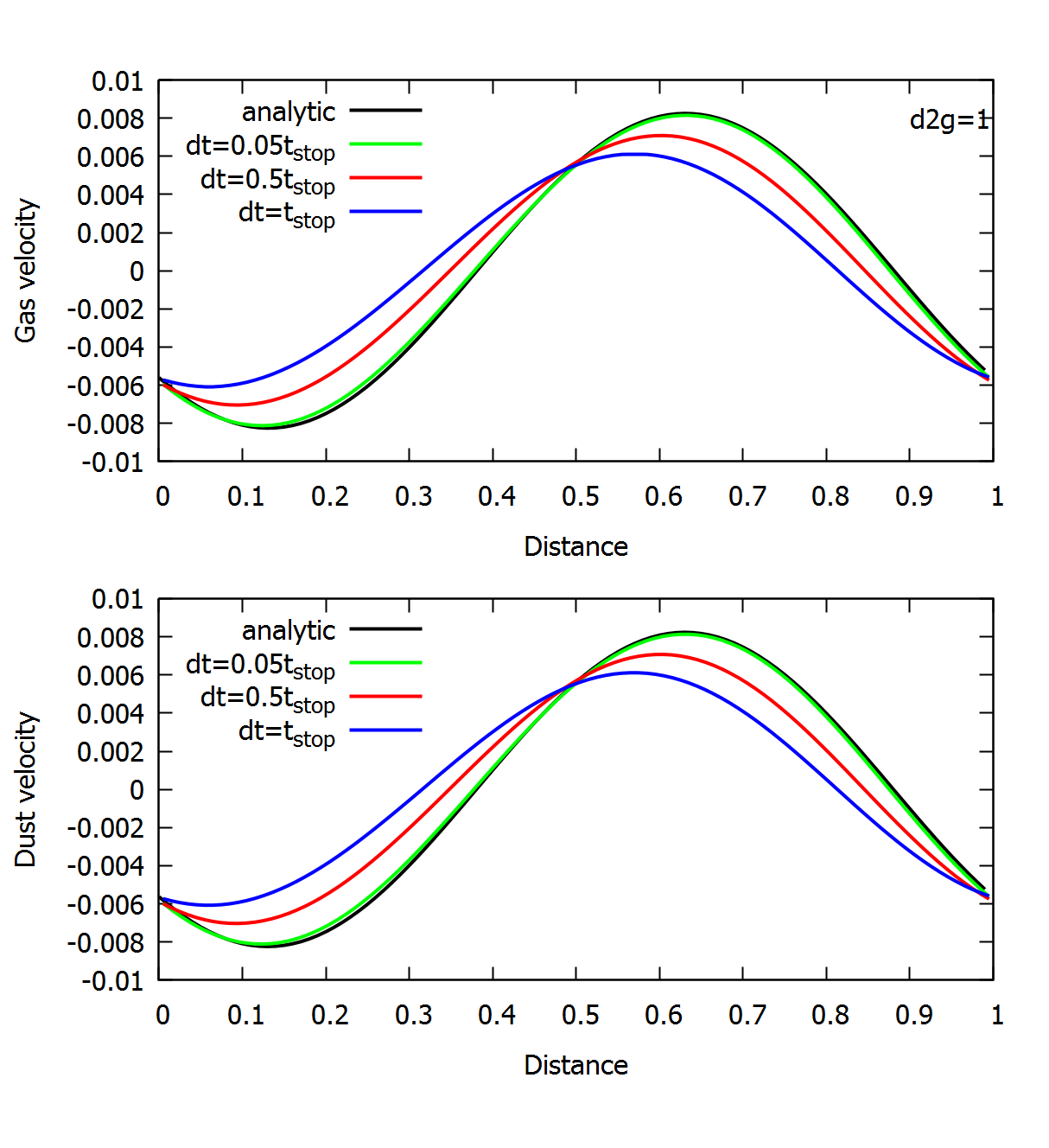}}
\par\end{centering}
\centering{}\protect\protect\protect\caption{\label{figA5} Solution of the dusty wave problem for 
strong coupling between gas and dust ($K=1000$) and the dust-to-gas ratio $d2g=1$ at time instance $t=0.5$. The top panel shows the gas velocity, while the bottom panel shows the dust velocity. The color lines present numerical solutions obtained with different time steps $dt$, while the black line shows the analytical solution.}
\end{figure}

\begin{figure}
\begin{centering}
\resizebox{\hsize}{!}{\includegraphics{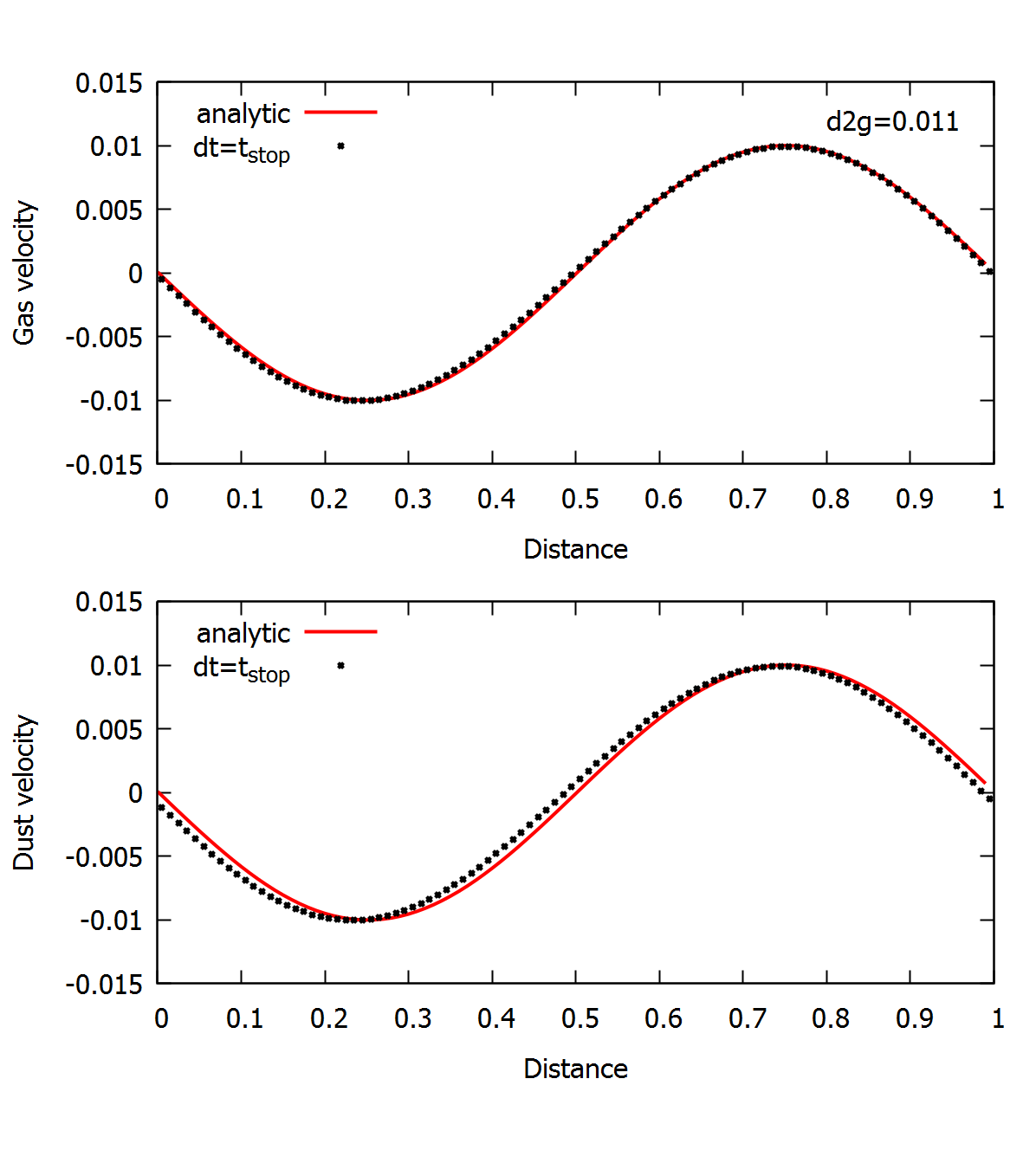}}
\par\end{centering}
\centering{}\protect\protect\protect\caption{\label{figA6} Solution of dusty wave problem for strong coupling between gas and dust ($K=100$) and the dust-to-gas ratio $d2g=0.011$ at time $t=0.5$. The top panel shows the gas velocity, while the bottom panel shows the dust velocity. The black dots present numerical solutions obtained with the time step $dt=t_{\rm stop}$, while the red line shows the analytical solution. }
\end{figure}

Figure~\ref{figA5} presents the gas and dust  velocities at $t=0.5$ for $K=1000$. The dust-to-gas ratio is $d2g=\tilde{\rho_{\rm d}} / \tilde{\rho_{\rm g}}=1.0$. As in Sect.~\ref{sec:ShockTube}, we 
varied the time step from $0.05 t_{\rm stop}$ to $t_{\rm stop}$. Clearly, 
%we see the same pattern as on the Figure~\ref{figA3}: for $dt=t_{\rm stop}$ 
the numerical solution differs notably from the analytical one for $dt=t_{\rm stop}$, but decreasing the time step to $dt=0.05 t_{\rm stop}$ enables to achieve a reasonably good agreement with 
the analytical solution for both the gas and dust velocities. 
We note that for smaller dust-to-gas ratios the requirement on the time step is less strict. 
Figure~\ref{figA6} presents the gas and dust velocities for the same time instance $t=0.5$ 
obtained for  $K=100$ and $d2g=0.011$.  
For $dt = t_{\rm stop}$, the agreement between the numerical and analytical solution is much better than for the case of $d2g=1.0$, but the dust velocity is calculated less accurately than the gas velocity. 

\begin{figure}
\begin{centering}
\resizebox{\hsize}{!}{\includegraphics{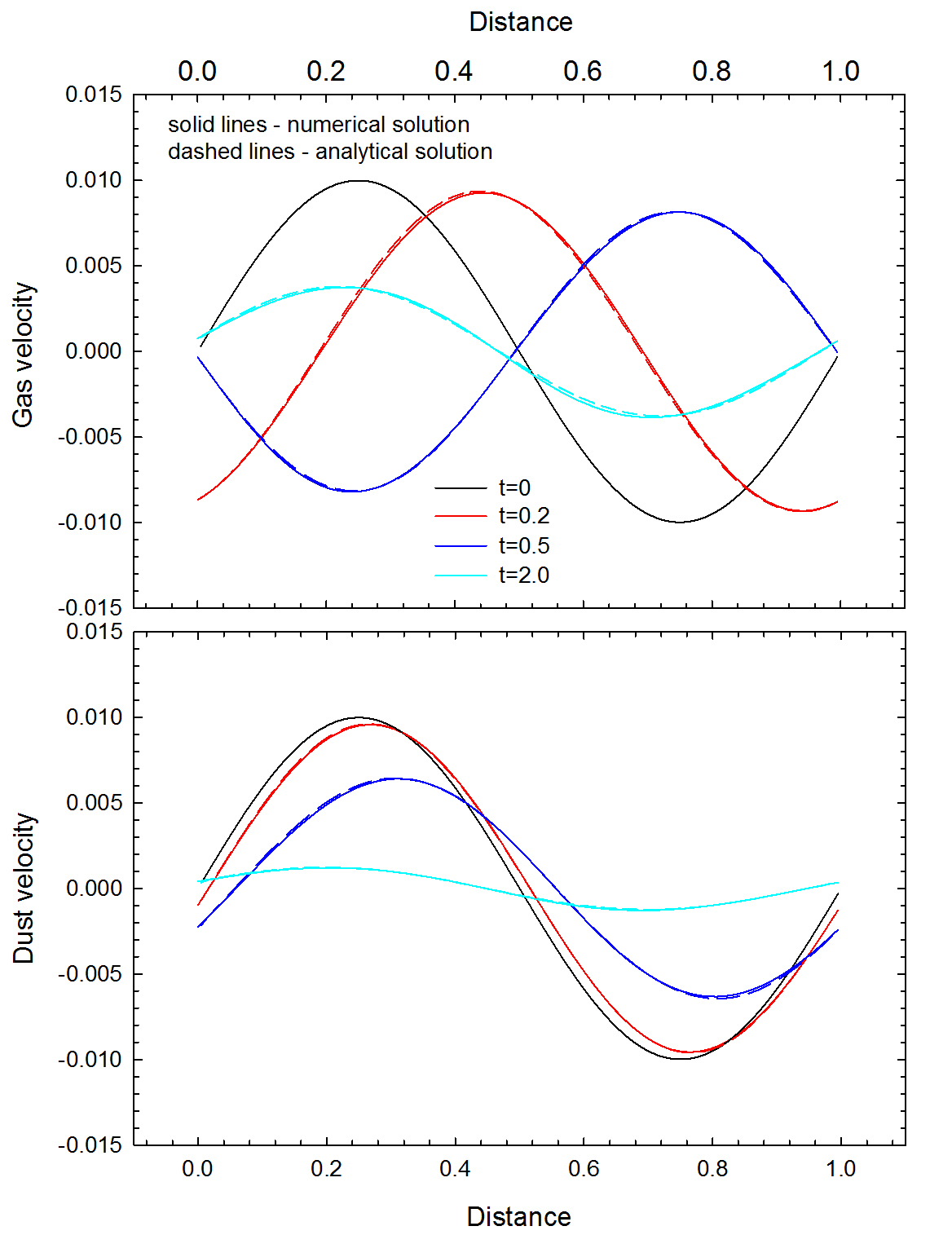}}
\par\end{centering}
\centering{}\protect\protect\protect\caption{\label{figA7} Solution of dusty wave problem for weak coupling between gas and dust ($K=1$) and the dust-to-gas ratio $d2g=1$ at times $t=0; \ 0.2; \ 0.5; \ 2.0$. The top panel shows the gas velocity, while the bottom panel shows the dust velocity. The solid lines present numerical solutions at different times as indicated in the legend, while the dashed lines present the corresponding analytical solutions. Note that both solutions are often almost indistinguishable.
}
\end{figure}

Finally, we consider the dusty wave with a weak coupling between the gas and dust, $K=1$. As was demonstrated
earlier, the most challenging for simulations is the case of comparable dust and gas densities. We therefore
choose $d2g=1.0$ for this test run.  The time step was defined by the Courant condition with a limiter $C=0.5$. The resulting gas and dust velocities are shown in Figure~\ref{figA7} for different time instances. Clearly, our scheme performs well for the case of weak coupling between the gas and dust even when back reaction is taken into account.

\end{appendix}

\end{document}